\documentclass{ucetd}
\usepackage{subfigure,epsfig,amsfonts}
\usepackage[square,sort,comma,numbers]{natbib}
\usepackage{amsmath}
\usepackage{amssymb}

\title{\texorpdfstring{$T{\bar T}$}{TTbar} and Holography}
\author{Asrat Demisē}
\department{Physics}
\division{Physical Sciences}
\degree{Doctor of Philosophy}
\date{August 2021}

\dedication{This dissertation is dedicated to my mother T'iru.}% $\d{T}$iru Tiruye Mekonnen
\epigraph{Epigraph Text}

\usepackage[pdfusetitle]{hyperref}
\hypersetup{unicode=true,
            linktoc=all,
            pdfsubject=subject here,
            pdfkeywords=keyword1 keyword2 keyword3,
            pdfborder={0 0 0},
            breaklinks=true}
\makeatletter
\let\ORG@hyper@linkstart\hyper@linkstart
\protected\def\hyper@linkstart#1#2{%
  \lowercase{\ORG@hyper@linkstart{#1}{#2}}}
\makeatother

\begin{document}
%% Basic setup commands
% If you don't want a title page comment out the next line and uncomment the line after it:
\maketitle
%\omittitle

% These lines can be commented out to disable the copyright/dedication/epigraph pages
\makecopyright
\makededication
%\makeepigraph

%% Make the various tables of contents
\tableofcontents
\listoffigures
%\listoftables

\acknowledgments
% Enter Acknowledgements here

It is a pleasure to acknowledge those who offered me their support, encouragement and guidance during my PhD
studies.

First and foremost, I would like to express my sincere gratitude to my advisor David Kutasov for his continued feedback and guidance over the years. I also thank him for his useful comments on the draft of the dissertation.

I would like also to thank Michael Levin, David Schmitz and Savdeep Sethi for serving on my dissertation committee. 

I thank Savdeep Sethi for discussions on projects related to my work and useful conversations. I also thank Jonah Kudler-Flam for inviting me to collaborate on a project and continued discussions on related topics.
 
 In the course of my PhD studies, I have met so many people who have supported, encouraged and taught me. I am grateful to all the people in the Enrico Fermi Institute and Physics Department. I am especially grateful to David Reid, Amy Schulz, Robert Wald and Paul Wiegmann.

 I am indebted to the people at ICTP from whom I learned so much about physics. I especially thank K. S. Narian who was my thesis advisor at ICTP.

Finally I would like to thank my friends in Hyde Park and outside and my family for their time, constant support and encouragement. I especially thank my mother. %I also would like to thank the community in Hyde Park, Chicago. The community is very welcoming and home of friendly people.

\abstract
% Enter Abstract here

In recent years, there have been two independent but related developments in the study of irrelevant deformations in two dimensional quantum field theories (QFTs). The first development is the deformation of a two dimensional QFT by the determinant of the energy momentum stress tensor, commonly referred to as $T{\bar T}$ deformation. The second development is in two dimensional holographic field theories which are dual to string theory in asymptotically Anti-de Sitter (AdS) spacetimes. In this latter development, the deformation is commonly referred to as single-trace $T{\bar T}$ deformation.

The single-trace $T{\bar T}$ deformation corresponds in the bulk to a string background that interpolates between AdS spacetime in the infrared (IR) and a linear dilaton spacetime (vacuum of little string theory (LST)) in the ultraviolet (UV). It serves as a useful tool and guide to better understand and explore holography in asymptotically AdS and non-AdS spacetimes in a controlled setting. In particular, it is useful to gain insights into holography in flat spacetimes. 

The dissertation is devoted to the study of single-trace $T{\bar T}$ deformation and its single-trace generalizations in theories with $U(1)$ currents, namely $J\bar T$ and $T\bar J$ deformations, in the context of gauge/gravity duality. In the dissertation I present new results in the study of holography in asymptotically non-AdS spacetimes. I discuss two point correlation functions in single-trace $T{\bar T}$ deformation, and entanglement entropy and entropic $c$-function in single-trace $T{\bar T}$, $J\bar T$ and $T\bar J$ deformations. 

I show that two point functions in position space have both real parts and imaginary parts. I also show that the imaginary parts are non-perturbative. The imaginary parts correspond in momentum space to branch cuts, which signal non-locality. I obtain exact result for entanglement entropy associated with a spatial region of finite size. I also show that in the UV for a particular combination of the deformation couplings the leading order dependence of the entanglement entropy on the size is given by a square root but not logarithmic function. Such power law dependence of the entanglement entropy on the size is quite distinct and interesting. I also give exact result for the entropic $c$-function and show that it is regularization schemes independent, positive and monotonic, which are similar to the behaviors observed in conventional local QFTs. I also discuss its distinctive features in the UV.

\mainmatter
% Main body of text follows

\chapter{Introduction}
% Introductory stuff

In this chapter we present a brief summary of recent new developments in the study of quantum field theories (QFTs) that are the foundations for the dissertation.  Along the way, we also discuss the motivations for the dissertation and its main focuses. We end the chapter with a brief discussion on the organization of the rest of the dissertation and on related published works that are not included in the subsequent chapters. The study of QFTs is generally the study of the different renormalization group (RG) flows. In what follows, we begin with a brief discussion on Wilsonian RG flow which relates a given quantum field theory (QFT) at different energy scales. 

\section{Wilsonian RG flow}

A quantum field theory (QFT) is normally defined by specifying a set of couplings and some cut-off scale. To study the theory at lower energy scales we apply  the renormalization group (RG) flow analysis. The RG is a group of scale transformations, and it is equivalent to a redefinition of the cut-off scale. Under the scale transformations, the couplings transform non-trivially. This corresponds in the coupling space of the theory a trajectory, often called flow, that connects the ultraviolet (UV) to the infrared (IR) regimes. Thus, the RG flow compares the theory at different energy scales.

The contemporary understanding of mathematically well-defined QFTs is based on fixed points of RG flow, and the Wilsonian idea is that a QFT is defined through an RG flow that originates from a fixed point. Fixed points represent scale invariant theories or conformal field theories (CFTs)\footnote{For a review on the distinctions and/or relations between scale invariant and conformal field theories see \cite{Nakayama:2013is}.}. The (Wilsonian) RG flow is determined by specifying a set of couplings at the fixed point. These couplings are associated with two classes of operators.

At a fixed point operators are in general grouped based on their (scaling) dimensions into three broad categories: irrelevant, relevant and marginal.

In general the number of relevant and marginal operators is finite, and these are the operators for which we need to specify the couplings. These couplings define the QFT and continue to describe all the possible RG flows to longer distances. The general understanding after the work of Wilson and others (see \cite{Wilson:1973jj}) is that along an RG flow in the IR the theory is well-defined and/or mathematically consistent and tractable. In general, the IR theory can be either non-trivial, that is, interacting or free, or empty, that is, it can have a mass gap. An example with a mass gap is non-abelian Yang-Mills (YM) theory in four dimensions. This theory is believed to develop a mass gap in the IR.%Examples are a theory with massless particles (for example, QED at energies below the rest mass of electron) and a theory with a mass gap (for example, non-abelian Yang-Mills (YM) theory in four dimensions).

We must mention here, however, the phenomenon that along an RG flow it may appear that an operator that was initially irrelevant (relevant) at the UV fixed point builds up negative (positive) anomalous dimension, and eventually becomes relevant (irrelevant) in the IR region. Such an operator that becomes relevant (irrelevant) in the IR is sometimes referred to as a dangerously irrelevant (harmlessly relevant) operator \cite{Asrat:2018dye}. Thus, in general one needs also to specify the couplings for the dangerously irrelevant operators since they are relevant away from the UV fixed point.

On the other hand, the number of irrelevant operators at a given fixed point is infinite. To study a QFT\footnote{The QFT is defined in the Wilsonian RG flow sense discussed above by deforming the fixed point by a set of (marginally) relevant operators.} at high energies we specify in the theory the couplings associated with the irrelevant operators. Thus, one has to consider a large number of irrelevant operators. This makes, in general, difficult to apply the RG analysis in a well organized and controlled way, and therefore, it limits the predictive power or domain of validity of the theory. Note also that in this case it is very likely that in the UV the theory may not be described by the original fundamental field degrees of freedom. In general we may, but not necessarily, attribute this to the existence of infinitely many irrelevant operators at the fixed point. In general, whether one is able to consider one or several of the infinitely many irrelevant operators, one will encounter singularities or ambiguities and/or will be led to introduce (the other) infinitely many couplings along the RG upflow towards the UV, and therefore, in general observables are not UV finite. For the above reasons, the general understanding has been, at least until recently, that RG upflows generated by irrelevant operators do not lead to well-defined and mathematically tractable quantum field theories.

\section{Recent developments}

In recent years, there have been two independent but related developments in the study of irrelevant deformations in two dimensional QFTs \cite{2004hep.th....1146Z, 2017NuPhB.915..363S, 2016JHEP...10..112C, 2017arXiv170105576G}. In these developments deformations generated by special operators are considered. Although the deformations are generated by irrelevant operators, several exact results are obtained and observables are UV finite. Thus, the UV theories, that is, the theories in the UV limit, are under better mathematical tractability. However, they are distinct from conventional local QFTs. In the dissertation, I will mainly study the second development which will be discussed shortly below. In particular, I study the UV theory in the corresponding irrelevant deformation using several observables as probes.%involve \footnote{This is in particular very clear in the second development, which we will discuss shortly below}along the RG upflows the deformations do not require introducing new couplings.  however,

The first development is the deformation of a two dimensional QFT by the determinant of the energy momentum stress tensor, commonly referred to as  $T{\bar T}$ deformation \cite{2004hep.th....1146Z, 2017NuPhB.915..363S, 2016JHEP...10..112C}. The generating $T{\bar T}$ operator is irrelevant and it is well-defined (in the coincidence limit up to total derivative terms). Since it is built out of (the components of) the stress tensor which is present and conserved in any QFT with spacetime translation invariance, the operator exists universally, and therefore also the deformation is universally applicable.  Given that the operator is well-defined in any QFT, it can be used to define a one parameter family of theories by iteratively adding it to and updating the Lagrangian density in a small increment of the coupling. This defines a trajectory in field theory space parametrized by the deformation coupling. Therefore, it can be used to probe and better understand the space of field theories.%it can be used to flow up the RG% , or equivalently, to flow up the RG,

In general, under an irrelevant deformation of a QFT, as we mentioned earlier, the understanding is that at high energies or in the UV the theory is ill-defined. The theory involves ambiguities and/or singularities and thus, it is not solvable. The $T{\bar T}$ deformation, however, defines a theory that is under better mathematical tractability and solvable. In particular, the deformation preserves some of the symmetries of the original theory, and in the resulting deformed theory, for several observables, exact and UV finite results are obtained. The results include energy spectrum, S-matrix and thermodynamic partition function \cite{2004hep.th....1146Z, 2017NuPhB.915..363S, 2016JHEP...10..112C, 2018JHEP...08..106D, 2018JHEP...10..186C, Asrat:2020jsh}. Other observables such as correlation functions are also obtained in various limits \cite{Cardy:2019qao, Aharony:2018vux, He:2019vzf}. At high energies the deformed theory exhibits Hagdorn density of states and non-locality. In particular, it is shown for a $T{\bar T}$ deformed CFT that there exists an energy above which the specific heat is negative \cite{Barbon:2020amo} and thus the deformed theory has a large density of states \cite{Aharony:1998tt}. These results indicate that at high energies the deformed theory is not governed by a conventional local fixed point. % away from the IR fixed point. flow equations

The deformation can be thought of as coupling the theory to a two dimensional gravity or a random geometry \cite{Dubovsky:2017cnj, Dubovsky:2018bmo, 2018JHEP...10..186C}. It can also be thought of as non-trivial field dependent transformations or redefinitions of coordinates and this, in particular, provides a simple explanation for the solvability of the deformed theory \cite{2018JHEP...10..186C, Conti:2018tca}. The deformation is also generalized in the presence of supersymmetry \cite{Chang:2018dge} and to other irrelevant deformations for theories with currents \cite{2018ScPP....5...48G}. These latter deformations involving $U(1)$ currents are commonly referred to as $J{\bar T}$ and $T{\bar J}$ deformations.

The second development is in two dimensional holographic CFTs which are dual to string theory in asymptotically AdS spacetimes \cite{2017arXiv170105576G}. In this case the deformation is commonly referred to as single-trace $T{\bar T}$ deformation. The single-trace $T{\bar T}$ operator which generates the deformation is irrelevant and present in any two dimensional holographic CFT. Thus, in two dimensional holographic CFTs the deformation is universally applicable. This deformation is closely related but distinct from the $T{\bar T}$ deformation\footnote{For a general discussion on single and multiple trace operators in string theory see \cite{Aharony:2001pa, Aharony:2001dp, Witten:2001ua}.}. The single-trace $T{\bar T}$ operator, as we will show in the next chapter, is given by a sum of $T{\bar T}$ operators and therefore the name single-trace. The $T{\bar T}$ deformation in this case is usually referred to as double-trace $T{\bar T}$ deformation for a reason that will be clear in the next chapter. In general, however, it is not known how to independently construct or define the single-trace $T{\bar T}$ operator in a generic two dimensional QFT that is not holographic and/or has no tensor (or symmetric) product structure. The existence of such single-trace operator in holography is related to a symmetric product structure of the quantum field theories. I will comment further on this in Chapter 5.

The single-trace $T{\bar T}$ deformation corresponds in the bulk to a string background that interpolates between $AdS_3$ in the IR and a linear dilaton spacetime $\mathbb{R}^{(1, 1)} \times \mathbb{R}$ in the UV. This background can be constructed by taking the zero string coupling limit in some configuration of fundamental strings and solitonic fivebranes. That is, the single-trace $T{\bar T}$ deformation has interpretation in terms of branes. String theory in the interpolating background is described by a deformed sigma model. The deformation corresponds to adding an exact marginal current bilinear operator to the IR Wess–Zumino–Witten (WZW) model. Thus, from the world-sheet perspective it is more evident why the dual RG upflow is under better mathematical tractability and that the dual deformed boundary theory is solvable. In the dual string theory the deformation is also equivalent to field dependent redefinitions or transformations of bulk coordinates or fields and therefore this further provides a simple explanation as to why the deformation in the boundary theory is solvable. The bulk description therefore provides a way to study non-local theories non-perturbatively in a controlled setting. Some quantities of interest such as energy spectrum, correlation functions, thermodynamic partition function and entanglement entropy are obtained \cite{Asrat:2017tzd, Giveon:2017myj, Chakraborty:2018kpr, Asrat:2019end, Hashimoto:2019wct}. 

At high energies the deformed boundary theory in this case also exhibits a Hagedorn density of states and non-locality. This deformation is also generalized to other irrelevant deformations for holographic theories with currents \cite{Chakraborty:2018vja}. These are commonly knows as single-trace $J{\bar T}$ and $T{\bar J}$ deformations.

The single-trace $T{\bar T}$ deformation and its generalizations are useful to better understand holography and gain insights into holography in flat spacetimes. In the dissertation I present new results in the study of holography in asymptotically non-AdS spacetimes. 

\section{Organization of the dissertation}

The dissertation is mainly devoted to the study of single-trace $T{\bar T}, J{\bar T}$ and $T{\bar J}$ deformations in the context of gauge/gravity duality. The organization of the dissertation is as follows.

In chapter two we introduce both $T{\bar T}$ and single-trace $T{\bar T}$ deformations in two dimensions. This chapter will be mainly a review.  We begin by defining the $T{\bar T}$ deformation. To demonstrate its relations to string theory or quantum gravity we consider the deformation in the theory of massless scalar fields\footnote{In the sense different from single-trace $T{\bar T}$ deformation. This will be clear in the next chapter.}. In particular, we derive the exact deformed Lagrangian density for a free massless scalar field at finite values of the deformation coupling. We also obtain the spectrum for a $T{\bar T}$ deformed QFT quantized on a circle. We discuss the main features of this result in the case the initial theory is a CFT. In particular, we discus the Casimir energy which defines in the deformed theory an effective central charge as a function of the size of the circle. We next explain the single-trace $T{\bar T}$ deformation in two dimensional holographic CFTs and its relations to little string theory (LST). We review the gauge/gravity duality and the NS5-NS1 branes configuration which in certain limits is related to vacuum of LST. We also demonstrate that in the dual string theory the single-trace $T{\bar T}$ deformation can also be interpreted as equivalent to momentum dependent spectral flow or field dependent redefinitions of bulk fields. We derive the single-trace $T{\bar T}$ deformed spectrum from the dual string theory and discuss its relations to the $T{\bar T}$ deformed spectrum. 

In chapter three we discuss correlation functions in the single-trace $T{\bar T}$ deformed boundary theory using the gauge/gravity duality. This chapter is based on the paper \cite{Asrat:2017tzd}. The main question that we address is concerning how the non-locality of the deformed theory is manifested in the analytic structure of correlation functions of operators that are local in the IR CFT. We in particular consider two point functions.

In chapter four we consider the general deformation involving the single-trace operators $T{\bar T}, J{\bar T}$ and $T{\bar J}$. This chapter is based on my paper \cite{Asrat:2019end}. We study entanglement entropy in the boundary theory associated to a spatial region of finite size from the dual string theory description. In holographic field theories, entanglement entropy is encoded in certain geometrical quantities in the bulk background, and the key observation of the paper is to note that there is an isometry of the dual bulk string background that preserves the boundary conditions imposed on the coordinates or fields. This observation is very crucial to obtain exact results. We also compute the Casin--Huerta entropic $c$--function. We study the monotonicity property of the entropic $c$--function along the RG upflow and its independence of regularization scheme that one introduces to regularize the UV divergence of entanglement entropy. This provides further support that the RG upflow is under better mathematical tractability.  It also gives important insight into the nature of the theory in the UV as it is not governed by a conventional local UV fixed point. We also briefly comment on the holographic proposals for the double-trace $T{\bar T}, J{\bar T}$ and $T{\bar J}$ deformations.

In chapter five we comment on how one may define the single-trace $T{\bar T}$ deformation in a generic QFT. As we mentioned earlier, unlike the $T{\bar T}$ deformation, the single-trace $T{\bar T}$ deformation is defined only for holographic theories or QFTs that have tensor product structures. In this chapter we put forward an idea that can be useful in defining the single-trace deformation in a generic QFT.

In the remainder of the current subsection I will briefly mention other related research projects that are published but not included in the dissertation.

In my paper \cite{Asrat:2020jsh} I studied the modular properties of Korteweg-De Vries (KdV) charges correlation functions in a $T{\bar T}$ deformed two dimensional CFT. A CFT has a symmetry algebra such as affine Lie algebra or Virasoro algebra. The universal covering algebra contains an abelian subalgebra generated by KdV charges. The $T{\bar T}$ deformation preserves the subalgebra. In the undeformed CFT the KdV charges have well-defined modular properties. I showed that this property also continues to exist in the resulting QFT after the deformation. I found that correlation functions decompose into a direct sum of two non-holomorphic but modular forms. I also obtained a general differential equation that the KdV generalized torus partition function obeys. The differential equation provides a non-perturbative description of $T{\bar T}$ deformed theories. I showed that the differential equation has a diffusion like interpretation with reaction terms which depend on spins.

In the collaborative paper \cite{Asrat:2020uib} we studied information theoretic quantities both in single and double-trace $T{\bar T}$ deformations at zero and finite temperatures. We computed mutual information and reflected entropy. For the single-trace deformation we found that the mutual information and reflected entropy diverge for disjoint intervals when the separation distance approaches a minimum finite value. This implies that the mutual information fails to serve as a geometric regulator which is related to the breakdown of the split property at the inverse Hagedorn temperature. In contrast, for the double-trace deformation we found all divergences to disappear including the standard quantum field theory ultraviolet divergence that is generically seen as disjoint intervals become adjacent. We furthermore computed reflected entropy in conformal perturbation theory. While we found formally similar behavior between bulk and boundary computations, we find quantitatively distinct results. We commented on the interpretation of these disagreements and the physics that must be altered to restore consistency. We also briefly discussed the $T{\bar J}$ and $J{\bar T}$ deformations.

%\texorpdfstring{$T\bar{T}$}{TEXT}

\chapter{\texorpdfstring{$T{\protect\bar T}$}{TTbar} deformation}

In this chapter we discuss both the $T{\bar T}$ deformation and the closely related single-trace $T{\bar T}$ deformation. We obtain the corresponding deformed spectrums. We begin with the $T{\bar T}$ deformation. We follow \cite{2004hep.th....1146Z, 2017NuPhB.915..363S, 2016JHEP...10..112C}. See also \cite{Jiang:2019hxb, GuicaCern20, Bonelli:2018kik}.

\section{\texorpdfstring{$T{\bar T}$}{TTbar} deformation}

We work in two dimensional Cartesian coordinates $(x, y)$. We define the complex variables
\begin{equation} \label{eq:2.2}
({\textrm{holomorphic}}) \ z = x + iy, \quad (\textrm{antiholomorphic}) \ {\bar z} = z^* = x - iy.
\end{equation}
We take $y$ as the Euclidean time. 

Consider a two dimensional quantum field theory (QFT)\footnote{The theory can have a gravity dual description in the context of gauge/gravity duality.} with energy momentum stress tensor ${T}_{\mu\nu}$. Define the composite operator ${{\textrm T}{\bar {\textrm T}}}$ by the determinant of ${T}_{\mu\nu}$ as
\begin{equation} \label{eq:2.111}
 {\textrm{det}} \left({T}_{\mu\nu}\right) := -{1\over \pi^2}{\textrm T}{\bar {\textrm T}}.
\end{equation}
The components of the energy momentum stress tensor are given in a particular normalization by
\begin{equation}\label{eq:2.3}
T := -2\pi T_{zz}, \quad {\bar T} := -2\pi T_{\bar{z}\bar{z}}, \quad \Theta := 2\pi T_{z{\bar z}}.
\end{equation}
With this normalization the composite operator \ref{eq:2.111} that generates the $T{\bar T}$ deformation takes the form\footnote{For a conformal field theory (CFT) the trace $\Theta$ is zero. Thus, ${{\textrm T}{\bar {\textrm T}}} = T{\bar T}$.}
\begin{equation}\label{eq:2.4z}
{{\textrm T}{\bar {\textrm T}}} = T{\bar T} - \Theta^2.%  =  \lim_{z\to z'} \left(T(z, {\bar z}){\bar T}(z', {\bar z'})-\Theta(z, {\bar z})\Theta(z', {\bar z'})\right).
\end{equation}

In the coincidence limit $z \to z'$,  as we will show shortly, the operators $T(z, {\bar z}){\bar T}(z', {\bar z'})$ and $\Theta(z, {\bar z})\Theta(z', {\bar z'})$ are both divergent but their non-derivative divergent parts cancel each other and therefore leaving the combination \ref{eq:2.4z} finite up to total derivative terms. Thus, the composite operator \ref{eq:2.4z} is well-defined (up to total derivative terms), and as a result it can be used to define a one parameter family of theories by iteratively adding it to and updating the Lagrangian density in a small increment of the deformation coupling. This defines a trajectory in field theory space. %, or equivalently, to flow up the RG,

We parametrize the trajectory by $t$ and denote the Lagrangian density at each point along the trajectory by ${\mathcal L}_t$. That is, each point $t$ along the trajectory represents a field theory described by a Lagrangian density ${\mathcal L}_t$. Along the trajectory the Lagrangian density obeys or flows according to the equation,
\begin{equation} \label{eq:2.1}
{d{\mathcal L}_{t}\over dt} =  {{\textrm{det}}} {T}^{(t)}_{\mu\nu} := -{1\over \pi^2}({\textrm T}{\bar {\textrm T}})^{(t)},
\end{equation}
where ${T}^{(t)}_{\mu\nu}$ is the energy momentum stress tensor of the theory at the point labeled by $t$ along the trajectory. The parameter $t$ serves as the deformation coupling. It has mass dimension $-2$. Thus, for instance, at $t = 0$ the composite operator is irrelevant in the Wilsonian RG flow sense. To leading order in $t$ the $T{\bar T}$ deformation is
\begin{equation} \label{eq:2.1lz}
{\mathcal L}_{t} =  {\cal L}_0 - {t\over \pi^2}({\textrm T}{\bar {\textrm T}})^{(0)} + {\cal O}(t^2).
\end{equation}

We next show that the composite operator \ref{eq:2.1} is well-defined. The composite operator \ref{eq:2.1} in terms of the components of ${T}^{(t)}_{\mu\nu}$ is given by
\begin{equation}\label{eq:2.4}
({{\textrm T}{\bar {\textrm T}}})^{(t)} = T^{(t)}{\bar T}^{(t)} - \left(\Theta^{(t)}\right)^2.
\end{equation}
To show that it is well-defined we need to make the following two assumptions.  We need to assume conservation of the energy momentum stress tensor, that is,  
\begin{equation}\label{eq:2.5}
{\bar\partial} T^{(t)} - \partial \Theta^{(t)} = 0, \quad \partial {\bar T}^{(t)} - {\bar \partial} \Theta^{(t)} = 0,
\end{equation}
and that for any two (local) operators ${\mathcal O}^{(t)}_{a}$ and ${\mathcal O}^{(t)}_{b}$ in the theory at the point $t$ along the trajectory the Operator Product Expansion (OPE) takes the following form 
\begin{equation}\label{eq:2.6}
{\mathcal O}^{(t)}_a(z_1, {\bar z_1}){\mathcal O}^{(t)}_b(z_2, {\bar z_2}) = \sum_c C_{ab}^c(z_1 - z_2, \bar{z}_1 - \bar{z}_2){\mathcal O}^{(t)}_c(z_2, {\bar z_2}).
\end{equation}
That is, besides the conservation equations \ref{eq:2.5}, we are also assuming that there exists a complete set of (local) operators at each point $t$ along the trajectory obeying the OPEs \ref{eq:2.6}. In general the OPE coefficients depend on the deformation coupling $t$ but for brevity we omit the explicit dependence on it. See \cite{Cardy:2019qao} for a discussion on how they evolve along the trajectory. Note that the coefficients are invariant under global (rigid) translation. 

We define the composite operator \ref{eq:2.4} by point splitting as
\begin{equation}\label{eq:2.7}
({{\textrm T}{\bar {\textrm T}}})^{(t)}(z, {\bar z}) = \lim_{z\to z'} \left(T^{(t)}(z, {\bar z}){\bar T}^{(t)}(z', {\bar z'})-\Theta^{(t)}(z, {\bar z})\Theta^{(t)}(z', {\bar z'})\right).
\end{equation}
Consider differentiating \ref{eq:2.7} with respect to $\bar z$. This gives 
\begin{eqnarray}\label{eq:2.8}
 \partial_{\bar z}\left(T^{(t)}(z, {\bar z}){\bar T}^{(t)}(w, {\bar w}) - \Theta^{(t)}(z, {\bar z})\Theta^{(t)}(w, {\bar w})\right) = & \cr
 \partial_{\bar z} T^{(t)}(z, {\bar z}) {\bar T}^{(t)}(w, {\bar w}) - \partial_{\bar z}\Theta^{(t)}(z, {\bar z}) \Theta^{(t)}(w, {\bar w}). &
\end{eqnarray}\iffalse
\begin{equation}\label{eq:2.8}
\partial_{\bar z}\left(T^{(t)}(z, {\bar z}){\bar T}^{(t)}(w, {\bar w}) - \Theta^{(t)}(z, {\bar z})\Theta^{(t)}(w, {\bar w})\right) = 
\partial_{\bar z} T^{(t)}(z, {\bar z}) {\bar T}^{(t)}(w, {\bar w}) - \partial_{\bar z}\Theta^{(t)}(z, {\bar z}) \Theta^{(t)}(w, {\bar w}).
\end{equation}\fi
Using the conservation equation \ref{eq:2.5} this becomes
\begin{eqnarray}\label{eq:2.9}
\partial_{\bar z}\left(T^{(t)}(z, {\bar z}){\bar T}^{(t)}(w, {\bar w}) - \Theta^{(t)}(z, {\bar z})\Theta^{(t)}(w, {\bar w})\right) = & \cr
 \partial_{ z} \Theta^{(t)}(z, {\bar z}) {\bar T}^{(t)}(w, {\bar w}) - \partial_{\bar z}\Theta^{(t)}(z, {\bar z}) \Theta^{(t)}(w, {\bar w}). &
\end{eqnarray}\iffalse
\begin{equation}\label{eq:2.9}
\partial_{\bar z}\left(T^{(t)}(z, {\bar z}){\bar T}^{(t)}(w, {\bar w}) - \Theta^{(t)}(z, {\bar z})\Theta^{(t)}(w, {\bar w})\right) =   \partial_{ z} \Theta^{(t)}(z, {\bar z}) {\bar T}^{(t)}(w, {\bar w}) - \partial_{\bar z}\Theta^{(t)}(z, {\bar z}) \Theta^{(t)}(w, {\bar w}). 
\end{equation}\fi
Adding the second equation from \ref{eq:2.5} after multiplying it by $\Theta^{(t)}$ to \ref{eq:2.9} and rearranging terms we get 
\begin{eqnarray}\label{eq:2.10}
 \partial_{\bar z}\left(T^{(t)}(z, {\bar z}){\bar T}^{(t)}(w, {\bar w}) - \Theta^{(t)}(z, {\bar z})\Theta^{(t)}(w, {\bar w})\right) = & (\partial_z + \partial_w)(\Theta^{(t)}(z, {\bar z}){\bar T}^{(t)}(w, {\bar w}))   \cr
  - & (\partial_{\bar z} + \partial_{\bar w})(\Theta^{(t)}(z, {\bar z})\Theta^{(t)}(w, {\bar w})).\cr 
\end{eqnarray}
Similarly, we have
\begin{eqnarray}\label{eq:2.11}
\partial_{ z}\left(T^{(t)}(z, {\bar z}){\bar T}^{(t)}(w, {\bar w}) - \Theta^{(t)}(z, {\bar z})\Theta^{(t)}(w, {\bar w})\right) =  & (\partial_z + \partial_w)(T^{(t)}(z, {\bar z}){\bar T}^{(t)}(w, {\bar w})) \cr
- & (\partial_{\bar z} + \partial_{\bar w})(T^{(t)}(z, {\bar z})\Theta^{(t)}(w, {\bar w})). \cr
\end{eqnarray}

Using the second assumption \ref{eq:2.6} in the OPEs on the right-hand sides of \ref{eq:2.10} and \ref{eq:2.11} we write\iffalse 
\begin{equation}\label{eq:2.12}
\partial_{\bar z}\left(T^{(t)}(z, {\bar z}){\bar T}^{(t)}(w, {\bar w}) - \Theta^{(t)}(z, {\bar z})\Theta^{(t)}(w, {\bar w})\right) = \sum_i A^i\partial_w {\mathcal O}^{(t)}_i(w, {\bar w}) + \sum_i B^i\partial_{\bar w} {\mathcal O}^{(t)}_i(w, {\bar w}),
\end{equation}\fi
\begin{eqnarray}\label{eq:2.12}
\partial_{\bar z}\left(T^{(t)}(z, {\bar z}){\bar T}^{(t)}(w, {\bar w}) - \Theta^{(t)}(z, {\bar z})\Theta^{(t)}(w, {\bar w})\right) =  & \sum_i A^i\partial_w {\mathcal O}^{(t)}_i(w, {\bar w}) \cr
 + & \sum_i B^i\partial_{\bar w} {\mathcal O}^{(t)}_i(w, {\bar w}),
\end{eqnarray}\iffalse
\begin{equation}\label{eq:2.13}
\partial_{ z}\left(T^{(t)}(z, {\bar z}){\bar T}^{(t)}(w, {\bar w}) - \Theta^{(t)}(z, {\bar z})\Theta^{(t)}(w, {\bar w})\right) = \sum_i C^i\partial_w {\mathcal O}^{(t)}_i(w, {\bar w}) + \sum_i D^i\partial_{\bar w} {\mathcal O}^{(t)}_i(w, {\bar w}),
\end{equation}\fi
\begin{eqnarray}\label{eq:2.13}
\partial_{ z}\left(T^{(t)}(z, {\bar z}){\bar T}^{(t)}(w, {\bar w}) - \Theta^{(t)}(z, {\bar z})\Theta^{(t)}(w, {\bar w})\right) =  & \sum_i C^i\partial_w {\mathcal O}^{(t)}_i(w, {\bar w}) \cr
 + & \sum_i D^i\partial_{\bar w} {\mathcal O}^{(t)}_i(w, {\bar w}),
\end{eqnarray}
for some functions $A^i, B^i, C^i, D^i$. We write also the OPE for the composite operator \ref{eq:2.4} as
\begin{equation}\label{eq:2.14}
T^{(t)}(z, {\bar z}){\bar T}^{(t)}(w, {\bar w}) - \Theta^{(t)}(z, {\bar z})\Theta^{(t)}(w, {\bar w}) = \sum_i F^i(z - w, {\bar z} - {\bar w}) {\mathcal O}^{(t)}_i(w, {\bar w}).
\end{equation}
for some functions $F^i$.

It follows from \ref{eq:2.12} and \ref{eq:2.13} that any operator ${\mathcal O}^{(t)}_i$ appearing in the expansion in \ref{eq:2.14} unless itself is a coordinate derivative of another (local) operator, comes with a constant (i.e. coordinate independent) coefficient $F^i$. In other words, the OPE \ref{eq:2.14} can be written as
\begin{equation}\label{eq:2.15}
T^{(t)}(z, {\bar z}){\bar T}^{(t)}(w, {\bar w}) - \Theta^{(t)}(z, {\bar z})\Theta^{(t)}(w, {\bar w}) = ({{\textrm T}{\bar {\textrm T}}})^{(t)}(w, {\bar w}) + {{\textrm{derivatives \ terms}}}.
\end{equation}
Thus, in the coincidence limit it is well-defined up to total derivates terms which will not be relevant unless the theory is defined on a spacetime with non-trivial topology.

\subsection{Free massless scalar field}

We now consider as an example the theory of a free massless scalar field $\phi$ in two dimensions and deform it with the $T{\bar T}$ operator. We show that this leads to the Nambu-Goto Lagrangian for a string in three dimensions. Therefore, in general the deformation is related to a quantum field theory coupled with gravity or random metrics.

The dynamics of the free field theory is governed by the local action functional 
\begin{equation}\label{eq:2.16}
S_{0} = \int d^2x\sqrt{g} {\mathcal L}_{0}, \quad {\mathcal L}_{0} = {1\over 2}X, \quad X = g^{\mu\nu}X_{\mu\nu}, \quad X_{\mu\nu} = \partial_\mu \phi \partial_{\nu} \phi,
\end{equation}
where $g_{\mu\nu}$ is an auxiliary metric tensor that we will use to compute the (Hilbert) energy momentum stress tensor. We replace the metric $g_{\mu\nu}$ with the flat spacetime metric at the end. The deformed Lagrangian density obeys the flow equation \ref{eq:2.1}
\begin{equation}\label{eq:2.17}
{d{\mathcal L}_{t}\over dt} =  {{\textrm{det}}} ({T}^{(t)}_{\mu\nu}) := -{1\over \pi^2}({\textrm T}{\bar {\textrm T}})^{(t)}.
\end{equation}
Lorentz invariance implies the deformed Lagrangian density ${\mathcal L}_{t}$ depends only on the deformation parameter $t$ and the Lorentz invariant scalar $X$. Therefore, we write ${\mathcal L}_{t} := {\mathcal L}(t, X)$. The energy momentum stress tensor is given by
\begin{equation}\label{eq:2.18}
T^{(t)}_{\mu\nu} = -{2\over \sqrt{g}}{\delta S_t\over \delta g^{\mu\nu}} = g_{\mu\nu}{\mathcal L}_t - 2{\partial {\mathcal L}_t\over \partial g^{\mu\nu}}, 
\end{equation}
and using this the ${T{\bar T}}$ composite operator is given by \ref{eq:2.4}
\begin{equation}\label{eq:2.19}
{\textrm{det}} ({T}^{(t)}_{\mu\nu}) = {\mathcal L}^2_t - 2 {\mathcal L}_t g^{\mu\nu}{\partial {\mathcal L}_t\over \partial g^{\mu\nu}} + 2\epsilon^{\mu\nu}\epsilon^{\rho\sigma} {\partial {\mathcal L}_t\over \partial g^{\mu\rho}}{\partial {\mathcal L}_t\over \partial g^{\nu\sigma}} ,
\end{equation}
where $\epsilon^{\mu\nu}$ is the two dimensional Levi-Civita symbol and satisfies the identity
 \begin{equation}\label{eq:2.20}
  \epsilon^{\mu\rho}\epsilon^{\nu\sigma} = g^{\mu\nu}g^{\rho\sigma} - g^{\rho\nu}g^{\mu\sigma}.
 \end{equation}
 
 After adding the $T{\bar T}$ operator the flow equation for the Lagrangian density becomes 
\begin{equation}\label{eq:2.21}
\partial_t {\mathcal L}_t + (X\partial_X - 1){\mathcal L}_t^2 = 0.
\end{equation}
The solution to this differential equation is given by
\begin{equation}\label{eq:2.22}
{\mathcal L}_t = -{1\over 2t} + {1\over 2t}\sqrt{1 + 4t{\mathcal L}_0} = -{1\over 2t} + {\mathcal L}_{NG},
\end{equation}
where ${\mathcal L}_{NG}$ is the Nambu-Goto string Lagrangian in the static gauge in three dimensions, and the coupling $t$ plays the role of $\alpha'$ which is the square of the string length $l_s$. In the Polyakov approach, the Nambu-Goto action can be simplified by introducing fluctuating random metrics (see \cite{Kiritsis:2007zza}). Thus, at the classical level we note that the $T{\bar T}$ deformed theory is related to string theory or a quantum theory coupled with gravity (or random metrics).

\subsection{Deformed spectrum}

We now obtain the energy spectrum for a $T{\bar T}$ deformed QFT quantized on a circle with circumference $R$. We make the identification $x \sim x + R$. Therefore, we are considering a $T{\bar T}$ deformed QFT on an infinite cylinder of size $R$. To obtain the spectrum we first consider the expectation value of the composite $({{\textrm T}{\bar {\textrm T}}})^{(t)}$ \ref{eq:2.4} operator in the state $|n; t\rangle$
\begin{equation}\label{eq:2.23}
C_n := \langle n; t|({{\textrm T}{\bar {\textrm T}}})^{(t)}|n; t\rangle = \langle T^{(t)}(z, {\bar z}){\bar T}^{(t)}(w, {\bar w})\rangle_n - \langle\Theta^{(t)}(z, {\bar z})\Theta^{(t)}(w, {\bar w})\rangle_n,
\end{equation}
where $|n; t\rangle$ is the eigenstate of the deformed Hamiltonian.%On the cylinder \ref{eq:2.15} is still valid since to derive it we only assumed translation (and rotational) symmetry or conservation of energy momentum tensor \ref{eq:2.5} and existence of OPEs \ref{eq:2.6}.

An important property of the object $C_n$ that is useful in obtaining the spectrum is its independence of the coordinates. We now show this property. We differentiate $C_n$ with respect to $\bar z$. This gives 
\begin{equation}\label{eq:2.24}
\partial_{\bar z}C_n =\langle\partial_{\bar z}T^{(t)}(z, {\bar z}){\bar T}^{(t)}(w, {\bar w})\rangle_n - \langle\partial_{\bar z}\Theta^{(t)}(z, {\bar z})\Theta^{(t)}(w, {\bar w})\rangle_n, 
\end{equation}
which using the conservation equation \ref{eq:2.5} becomes
\begin{equation}\label{eq:2.25}
\partial_{\bar z}C_n = \langle\partial_{z}\Theta^{(t)}(z, {\bar z}){\bar T}^{(t)}(w, {\bar w})\rangle_n - \langle\partial_{\bar z}\Theta^{(t)}(z, {\bar z})\Theta^{(t)}(w, {\bar w})\rangle_n. 
\end{equation}
We assume that for any (local) field ${\mathcal O}_i$ the expectation value $\langle{\mathcal O}_i\rangle_n$ is a constant independent of $z, {\bar z}$. We also assume global translation invariance of OPE coefficients or \ref{eq:2.6}. These lead, making use of the second equation of \ref{eq:2.5}, to
\begin{equation}\label{eq:2.26}
\partial_{\bar z}C_n = -\langle \Theta^{(t)}(z, {\bar z})\partial_{w}{\bar T}^{(t)}(w, {\bar w})\rangle_n + \langle\Theta^{(t)}(z, {\bar z})\partial_{ w}{\bar T}^{(t)}(w, {\bar w})\rangle_n = 0.
\end{equation}
Similarly, we find 
\begin{equation}\label{eq:2.27}
\partial_{ z}C_n = 0,
\end{equation}
Therefore, the auxiliary object defined in \ref{eq:2.23} is independent of the coordinates.
 
Another important property of $C_n$ is its factorization property. We next show that $C_n$ or the expectation value of $T{\bar T}$ factorizes into expectation values of the components of the energy momentum stress tensor. We first note that the two point function decomposes by inserting the identity as
\begin{eqnarray}\label{eq:2.28}
\langle T^{(t)}(z, {\bar z}){\bar T}^{(t)}(z', {\bar z}')\rangle_n =  & \cr
  \sum_{n'}  \langle n; t|T^{(t)}(z, {\bar z})|n'; t\rangle \langle n'; t|{\bar T}^{(t)}(z, {\bar z})|n; t\rangle e^{({\mathcal E}_n - {\mathcal E}_{n'})|y - y'|}e^{i(P_n - P_{n'})(x - x')}. &
\end{eqnarray}\iffalse
\begin{equation}\label{eq:2.28}
\langle T^{(t)}(z, {\bar z}){\bar T}^{(t)}(z', {\bar z}')\rangle_n = \sum_{n'}  \langle n; t|T^{(t)}(z, {\bar z})|n'; t\rangle \langle n'; t|{\bar T}^{(t)}(z, {\bar z})|n; t\rangle e^{({\mathcal E}_n - {\mathcal E}_{n'})|y - y'|}e^{i(P_n - P_{n'})(x - x')}.
\end{equation}\fi
We also find a similar decomposition for $\langle\Theta^{(t)}(z, {\bar z})\Theta^{(t)}(w, {\bar w})\rangle_n$. Therefore, for the combination in $C_n$ \ref{eq:2.23} to be independent of the coordinates, all terms in these decompositions with $n'\neq n$ must cancel out. We assume the spectrum is non-degenerate\footnote{This assumption is not necessary if we also consider the KdV charges.}. Thus, we have the following factorization formula
\begin{equation}\label{eq:2.29}
C_n = \langle n; t|{\textrm{det}} (T^{(t)}_{\mu\nu}) |n; t\rangle = \langle n; t|T^{(t)}|n; t\rangle \langle n; t|{\bar T}^{(t)}|n; t\rangle - \langle n; t|\Theta|^{(t)}n; t\rangle \langle n; t|\Theta^{(t)}|n; t\rangle.
\end{equation}
We mention here that for a QFT on a curved spacetime the factorization formula does not hold since the metric is dynamical. In a $T{\bar T}$ deformed CFT on a curved spacetime it holds only to leading order in the large central charge limit \cite{Jiang:2019tcq}.

From the definition of the energy-momentum tensor we have 
\begin{equation}\label{eq:2.30}
\langle n; t |T^{(t)}_{yy}|n; t\rangle = -{1\over R}{\mathcal E}_n ({\textrm{energy \ density}}),
\end{equation}
\begin{equation}\label{eq:2.31}
 -\langle n; t |T^{(t)}_{xx}|n; t\rangle = {d\over dR}{\mathcal E}_n ({\textrm{pressure}}), 
 \end{equation}
 \begin{equation}\label{eq:2.32}
 -i\langle n; t |T^{(t)}_{xy}|n; t\rangle = {1\over R}P_n (\textrm{momentum \ density}).
\end{equation}
Since the theory is quantized on a circle, the momentum is quantized in units of the size of the circle. That is,
\begin{equation}\label{eq:2.321}
\partial_t P_n = 0.
\end{equation}

Making use of the factorization formula \ref{eq:2.29} and the flow equation in the Hamiltonian formulation\footnote{The Hamiltonian is similar to the Lagrangian in Euclidean spacetime.} which is
\begin{equation}\label{eq:2.33}
{1\over R}\partial_t {\mathcal E}_n = -\langle n; t|{\textrm{det}}( T^{(t)}_{\mu\nu})|n; t\rangle,
\end{equation}
we find the equation
\begin{equation}\label{eq:2.331}
\partial_t{\mathcal E}_n + {\mathcal E}_n\partial_R{\mathcal E}_n = - {P_n^2\over R}. 
\end{equation}
This partial differential equation (PDE) is related to the inviscid Burgers equation that appear in the study of the theory of turbulence in fluid mechanics \cite{BURGERS1948171}\footnote{ In the case in which $p_n = 0$ the PDE is the inviscid Burgers equation.}.

In the case in which the original theory is a CFT the PDE \ref{eq:2.331} can be solved exactly and the spectrum takes a simpler form. Solving the equation \ref{eq:2.331} with the assumption that $t \geq 0$ and that at $t = 0$ the theory is a CFT gives the spectrum 
\begin{equation}\label{eq:2.34}
{\mathcal E}_n(t) = {R\over 2t}\left (\sqrt{1 + {4t E_n\over R} + {4t^2 P_n^2\over R^2}} - 1\right ), \quad {\mathcal E}_n(t = 0) = E_n. 
\end{equation}
We note that the energy of a state in the deformed theory depends only on the energy and momentum of the corresponding state in the undeformed CFT. We also note that as a consequence of the factorization formula \ref{eq:2.29} there is no mixing of states. Therefore, along the trajectory we are not creating or introducing new states. The deformed energies ${\mathcal E}_n$ are in general positive (and real) provided the coupling $t$ and the energies $E_n$ are positive (and real). We note that states with energies
\begin{equation}\label{eq:2.35}
E_n = P_n,
\end{equation}
do not deform. That is, the energies do not flow ${\mathcal E}_n = E_n$. We also note that in general in the case in which $t \gg R^2$ the spectrum is independent of the size of the circle, and this suggests that in the UV the theory is non-local (see \cite{Cardy:2019qao}).

\subsection{Casimir energy}

\begin{figure}[h!]
\begin{center}
\includegraphics[width=12cm,height=7.5cm]{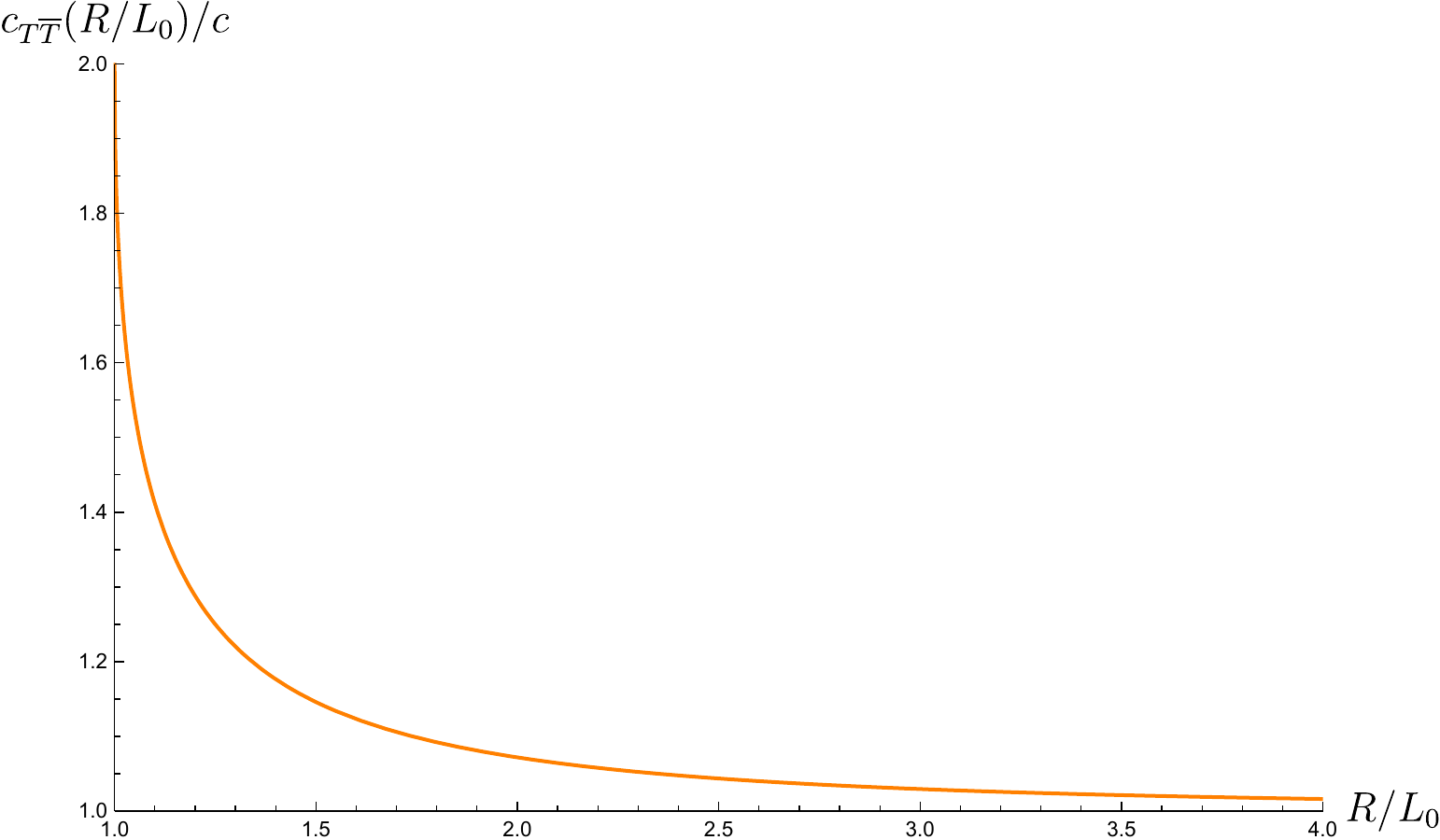}
\caption{The plot shows the normalized effective central charge $c_{T{\bar T}}$ as a function of the size of the circle $R$ per the minimum size $L_0$. %The range of the coupling $\lambda$ is between $0$ and ${6/\pi c}$.
}
\label{fig:2}
\end{center}
\end{figure}

In this section we discuss the deformed finite-size ground state energy or Casimir energy of a CFT. For a CFT we have
\begin{equation}\label{eq:2.36}
E_n(R) = {2\pi \over R}\left( h_n + {\bar h}_n - {c\over 12}\right), \quad P_n = {2\pi\left(h_n - {\bar h}_n\right)\over R},
\end{equation}
where $h_n$ and $\bar h_n$ are the eigenvalues of the Virasoro generators $L_0$ and $\bar L_0$, and $c$ is the central charge of the theory. The ground state or Casimir energy of the deformed theory is given by 
\begin{equation}\label{eq:2.37}
{\mathcal E}_0 = {1\over \pi \bar R \lambda}\left(\sqrt{1 + 2\lambda \pi \bar R E_0} - 1\right) = {1\over \pi \bar R \lambda}\left(\sqrt{1 - {c\lambda \pi\over 6}} - 1\right) = -{c_{T{\bar T}}(\lambda)\over 12 \bar R},
\end{equation}
where the dimensionless deformation parameter $\lambda$ is defined as
\begin{equation}\label{eq:2.38}
\quad \lambda = {t\over \pi^2 \bar R^2}, \quad \bar R = {R\over 2\pi},
\end{equation}
and the $c$-number $c_{T{\bar T}}(\lambda)$ defines an effective central charge \cite{Bloete:1986qm, Cardy:1986ie, Ludwig:1987gs, Cardy:1987gsx}
\begin{equation}\label{eq:2.388x}
c_{T{\bar T}}(\lambda) = {12\over \pi \lambda}\left(1 - \sqrt{1 - {c\lambda \pi \over 6}}\right), \quad c = c_{T{\bar T}}(\lambda = 0).
\end{equation}
We assume $\lambda$ is positive. We note that the deformed Casimir energy $\propto -c_{T{\bar T}}(\lambda)$ becomes complex unless the values of the dimensionless coupling $\lambda$ are restricted to lie in the interval
\begin{equation}\label{eq:2.40}
0\leq \lambda \leq {6\over \pi c}.
\end{equation}
Therefore, the Casimir energy becomes complex if the circumference or perimeter of the circle is  
\begin{equation}\label{eq:2.39a}
 2\pi \bar R <  L_0 :=  2\sqrt{\pi  ct\over 6}.
\end{equation}
Thus, $T{\bar T}$ deformation introduces a minimum size or distance $L_0 \propto \sqrt{ct}$. In terms of the minimum size $L_0$ and the size of the circle $R$ the effective central charge $c_{T{\bar T}}$ takes the form
\begin{equation}\label{eq:2.39}
{c_{T{\bar T}}(R)\over c} = 2\cdot {R^2\over L_0^2}\left(1 - \sqrt{1 - {L_0^2 \over R^2}}\right).
\end{equation}
In the case the size of the circle $R < L_0$ the effective central charge $c_{T{\bar T}}$ is complex and it has a positive real part and a negative imaginary part. The interpretation of the imaginary part is not completely clear\footnote{Purely imaginary central charges in general signal non-unitarity. In $T{\bar T}$ however the effective central charge has a positive real part in addition to an imaginary part. The real part is zero in the case $R < L_0$ only if the imaginary part is zero, and thus we do not expect a violation of unitarity.}. See Figure \ref{fig:2} for a plot of the effective central charge $c_{T\bar T}$ as a function of the size of the circle $R$. %Therefore, a $T{\bar T}$ deformed CFT cannot be put or is not defined on a cylinder of size $R$ less than the minimum distance $L_0$. 
% In the case in which $R < L_0$ the effective central charge $c_{T{\bar T}}$ develops an imaginary part. 

Note that as we take the central charge $c$ large the range \ref{eq:2.40} of admissible values of the coupling $\lambda$ shrinks more and more.

The existence of the minimum distance $L_0$ implies a maximum (also called Hagedorn) temperature $T_H = 1/\beta_H$\footnote{ The Hagedorn temperature was first introduced in the thermodynamic studies of hadrons \cite{Hagedorn:1965st}.}, and $\beta_H \propto L_0$. To show this relationship we consider the high energy limit of \ref{eq:2.34}. Since (we are assuming that) the energy eigenstates are non-degenerate, we can use Cardy's asymptotic entropy formula \cite{Cardy:1986ie}. The formula gives the high temperature behavior of the entropy. We find %This is a UV/IR relation.
\begin{equation}\label{eq:2.401}
S = 2\cdot 2\pi\sqrt{{c\over 6} \cdot {E\bar R\over 2}} \approx  4\pi\sqrt{{c\over 6} \cdot {t\over 4 \pi}}{\mathcal E} = L_0{\mathcal E} = \beta_H {\mathcal E}.
\end{equation} 
where $E$ is the total energy of the theory at $t = 0$. We see a related result in chapter four. %However, the Hagedorn temperature is not related to a phase transition. Therefore, $L_0$ sets the non-locality scale of the theory.  

\section{Single trace \texorpdfstring{$T{\bar T}$}{TTbar} deformation}

In this section we discuss the single-trace $T{\bar T}$ deformation and its close relations with the double-trace $T{\bar T}$ deformation. We obtain the single-trace $T{\bar T}$ deformed spectrum. We begin with a general and brief review on gauge/gravity duality. %We discuss correlation functions in the next chapter.

\subsection{The Gauge/Gravity duality}

The gauge/gravity duality or as often called the AdS/CFT correspondence is a relationship between field theory and gravity \cite{Maldacena:1997re, Gubser:1998bc, Witten:1998qj, Aharony:1999ti, Gubser:1997yh}. The general idea of the relationship is that string theory or M theory on some background is related to and the same as some field theory. The relationship is a duality because when the string theory is weakly coupled the field theory is strongly coupled and vice versa. Historically, there are two motivations and/or rationales for the relationship.% when it was first conjectured. \footnote{The relationship is a conjecture, and there are strong evidences to believe that it is correct.}

One motivation comes from 't Hooft planar limit of certain (gauge) field theories \cite{tHooft:1973alw}. The limit involves taking the rank of the gauge group (or the number of fundamental field degrees of freedom) to infinity while keeping the (relevant)  't Hooft coupling fixed. In this limit only planar Feynman diagrams dominate scattering amplitudes and non-planar Feynman diagrams are suppressed by the inverse of the rank of the gauge group with some exponents. This looks similar to the loop (or genus) expansion of scattering amplitudes in string theory after identifying the string coupling with the inverse of the rank of the gauge group. These string amplitudes are also related to the amplitudes that describe the dynamics of the (hadronic) strings that represent the flux tubes of gluon fields connecting quarks.

The other motivation comes from two different views or descriptions of branes in string theory \cite{Horowitz:1991cd, Duff:1993ye, Polchinski:1995mt, Stelle:1996tz, Duff:1996zn}. One description treats them as non-perturbative (solitonic) solutions of supergravity (to be more precise string or M theory), in particular, as black holes with throat or near horizon geometries. The supergravity backgrounds can be thought of as created by the branes and matter. The other description treats the branes as D(irichlet)-branes. D-branes are extended objects in spacetime where the endpoints of open strings reside. They cary certain charges and have tensions. Vector gauge and matter fields arise from strings that end on the branes. Thus, these fields are confined on the branes world-volume. They, however, interact with bulk fields such as graviton. %the full 

The two views are different descriptions of the same object and lead in the low energy limit, where we keep the (relevant) coupling on the D-branes world-volume fixed and send the Regge slope $\alpha'$ and/or Planck length $l_p$ to zero, to the correspondence or equivalence between the throat region with a given boundary conditions and the non-gravitational field theory on the D-branes world-volume. In this limit both the bulk and the Minkowski spacetime far away from the throat are similar and decouple.

The simplest case of the correspondence is the equivalence between type IIB string theory on $AdS_5\times S^5$ and ${\mathcal N} = 4$ four dimensional $U(N)$ SYM. The theory on the gravity side has two parameters. These are the radius $R$ in string units $l_{s}$ or Planck units $l_p$ and the string coupling $ g_{s}$. These parameters are related on the field theory side to the rank $N$ of the SYM gauge group and the SYM coupling $ g$. The relations are
\begin{equation}\label{eq:2.41}
{\alpha'\over R^2} \sim {1\over \sqrt{\lambda}}, \quad {\kappa^2_{10}\over R^8} \sim {1\over N^2}, \quad \alpha' = l_s^2, \quad \lambda = g^2N, \quad g^2 \sim g_{s},
\end{equation}
where $\kappa_{10} = \sqrt{8\pi G_{10}} \sim l_p^4$ is the ten dimensional gravitational constant and $\lambda$ is the 't Hooft coupling. 

We note that $1/N$ corrections in gauge theory become quantum loop effects in supergravity, while expanding in strong ('t Hooft) coupling amounts to $\alpha'$ corrections to supergravity. The gauge/supergravity theory correspondence is reliable only when supergravity is valid. That is
\begin{equation}\label{eq:2.42}
R \gg l_{s} \implies g\sqrt{N} \gg 1, \quad R \gg  l_p  \implies N \gg 1,
\end{equation}
so that the geometry is smooth. The correspondence is conjectured to be true for all values of $\lambda$ and $N$.

The low energy supergravity solution corresponding to a type II Ramond-Ramond (RR) charged black $p$-brane has in general a near horizon AdS geometry and the correspondence is the statement that supergravity or to be more precise string theory on a $d + 1$ dimensional AdS spacetime times a compact manifold with a given boundary conditions is dual to a $d$ dimensional conformal field theory. AdS spaces are maximally symmetric solutions of the Einstein equations with a negative cosmological constant. The isometry group $SO(2, d)$ of the bulk AdS spacetime is the group of conformal symmetries of the one dimension lower boundary. Therefore, it is useful to think of the conformal field theory as living on the boundary of the AdS spacetime. The boundary is at a finite conformal spatial distance and a light ray can reach the boundary in a finite time. Therefore, AdS spacetime is a special box. The correspondence is also holographic since the field theory is defined in at least one lower spacetime dimension. %which for convenience and naturally is taken to live on the boundary of the near horizon AdS geometry. Anti-de Sitter

In the conformal field theory the natural/good observables are not S-matrices as in QFT on flat spacetime but rather correlation functions of local operators inserted on arbitrary boundary points emitting and absorbing particles that propagate or are propagating in the interior (for related discussions see \cite{Alday:2007hr, Fitzpatrick:2011jn, Giddings:1999qu}). In the correspondence, we identify generating functional for correlation functions of local operators ${\mathcal O}^I$ in the conformal field theory with the partition function of string or M theory with some given boundary values for the fields $\Phi^I$,
\begin{equation}\label{eq:2.43}
\langle e^{\int{\mathcal L} +  \Phi_0^I{\mathcal O}^I}\rangle = \langle \left.e^{W}\right|_{\Phi^I\to \Phi^I_0 U^{\lambda_I}}\rangle_{ds^2},\quad ds^2 \propto {dU^2\over U^2} + U^2 dx^2,
\end{equation}
where $\lambda_I$ is determined by the mass of the field $\Phi^I$ which act on the boundary as a source for the operator ${\mathcal O}^I$. The coordinate $U$ labels the radial direction, and $\Phi^I_0$ is the boundary value of the field $\Phi^I$ and it completely determines the solution to the field equation of $\Phi^I$ in the AdS spacetime $ds^2$. %D-branes boundary

We next consider a brane configuration that is relevant for our discussion on single-trace $T{\bar T}$ deformation.

\subsection{NS5 - NS1 Branes System} 

S-duality\footnote{S-duality interchanges D1 with NS1, D5 with NS5, and it leaves D3 invariant \cite{Skenderis:1999bs}.} or electric/magnetic duality requires the existence of NS5 brane \cite{Vafa:1997pm}. NS5 is a 5-brane that is the magnetic dual of the elementary or fundamental string that we denote here by NS1. NS1 couples with the Neveu-Schwarz (NS) two-form. In type II string theories both the string and fivebrane preserve half the supersymmetries. % of the low energy supergravity. 

The low energy type II supergravity solution corresponding to $k$ coincident NS5 branes in the extremal limit is given by \cite{Callan:1991at, Duff:1995yh, Skenderis:1999bs}
\begin{equation}\label{eq:2.44}
ds^2 = (-dt^2 + dx_1^2\cdots + dx_5^2) + H_5(dr^2 + r^2 d\Omega^2_{3}),
\end{equation}
\begin{equation}\label{eq:2.45}
e^{2\Phi} = g^2_s H_5, \quad F_3 = \star dH_5, \quad H_5 = 1 + {R_5^2\over r^{2}}, \quad {R_5^2\over \alpha'} = k.
\end{equation}
The metric $ds^2$ is given in the string frame. The coordinates $t, x_1, x_2, x_3, x_4, x_5$ label the directions along the fivebranes and $d\Omega^2_{3}$ is the metric on the unit three dimensional sphere $S^3$ in the space normal to the fivebranes. $\Phi$ is the dilaton field and $g_s$ is the asymptotic value of the string coupling. The Hodge dual is taken in the transverse space and $F_3$ is the Neveu-Schwarz--Neveu-Schwarz (NSNS) 3-form flux. % 

We now consider taking the decoupling limit where we take the string coupling $g_s$ and Planck length $l_p$ to zero. Note that we are not taking $\alpha'$ to zero so that the field theory coupling on the world-volume of the branes is finite\footnote{On the NS5 branes the coupling $g^2 \sim \alpha'$.}. In this limit we obtain for the metric and dilaton 
\begin{equation}\label{eq:2.46}
ds^2 = (-dt^2 + dx_1^2 + \cdots + dx_5^2 + d\phi^2) + R_5^2 d\Omega_3^2, 
\end{equation}
\begin{equation}\label{eq:2.47}
 \Phi = -{1\over R_5}\phi.
\end{equation}
The boundary is at large positive $\phi$. We note that the string coupling vanishes for large positive $\phi$ and it diverges as one approaches to large negative $\phi$. Thus, studying string theory in the large negative $\phi$ region requires understanding the strong coupling behavior of the theory. 

String theory in the linear dilaton background $\mathbb{R}^{5, 1}\times \mathbb{R}_{\phi} \times S^3$  \ref{eq:2.46} is conjectured to be holographically dual to a six dimensional non-gravitionional theory on the fivebranes world-volume. The six dimensional theory is both similar to and different from the ten dimensional string theories. It is similar in that it describes interacting strings. The strings have a string length proportional to $l_s$ (see for example \cite{Banks:1999az, Harmark:2000qm}). It also exhibits Hagedorn spectrum and duality symmetries. However, it is different because it is six-dimensional and contains no graviton. For these reasons it is called little string theory (LST) \cite{Losev:1997hx, Aharony:1999ks, Kutasov:2001uf}. The LST can be defined with either type IIA or type IIB fivebranes (see \cite{Barbon:2007za}).

We now consider adding $p$ NS1 strings along one of the coincident NS5 branes spatial directions. We choose here the direction $x_1$. We compactify the remaining spatial directions of the NS5 branes on a compact four dimensional manifold ${\mathcal N}$. The simplest examples are the Calabi-Yau manifolds $T^4$ (which breaks no supercharges) and $K_3$ (which breaks half of the supercharges).  The full background takes in this case the structure ${\mathcal M}_3 \times {\mathcal N}\times S^3$. The relevant metric and dilaton field then become  \cite{Callan:1991at, Duff:1995yh, Skenderis:1999bs} 
\begin{equation}\label{eq:2.48}
ds^2 = H_1^{-1}(-dt^2 + dx^2_1) + R_5^2  d\phi^2,
\end{equation}
\begin{equation}\label{eq:2.49}
e^{2\Phi} = R_5^2 e^{-2\phi}H_1^{-1}, \quad H_1 = 1 + R_1^2e^{-2\phi}, \quad {R_1^2\over \alpha'} = {p\over \nu},
\end{equation}
where $\nu$ is proportional to the volume of the compact manifold $\mathcal N$ in units of $\alpha'^2$. 

We note that for large negative $\phi$ the (six dimensional) string coupling is $g_6 = R_5/R_1$ (and the three dimensional string coupling is $g_3^2 = 1/ (p\sqrt{k})$) which can be made small by taking $p$ (and $k$) large. Therefore we can apply perturbative string techniques (or free field representations) to study physics in this region. We also note that in this region the background ${\mathcal M}_3$ (\ref{eq:2.48}) is $AdS_3$ (with radius $R_5$ after shifting the $\phi$) which is dual to a two dimensional conformal field theory. For large positive $\phi$ the string coupling goes to zero and the background ${\mathcal M}_3$ (\ref{eq:2.48}) is the three dimensional linear dilaton background. The linear dialton background is holographically dual to a two dimensional compactification of LST.

Thus, the string background ${\mathcal M}_3$ (\ref{eq:2.48}) interpolates between $AdS_3$ in the IR or large negative $\phi$ and the three dimensional linear dilaton background $\mathbb{R}^{1, 1}\times \mathbb{R}_{\phi}$ in the UV or large positive $\phi$. We next show that the background ${\mathcal M}_3$ corresponds to an exact marginal current bilinear deformation of the world-sheet string theory on $AdS_3$. This defines in the boundary theory the single-trace $T{\bar T}$ deformation \cite{2017arXiv170105576G}.

The world-sheet string theory on the background $AdS_3$\footnote{The string background we obtained in the large negative $\phi$ limit of ${\mathcal M}_3$ (\ref{eq:2.48}).} supported with the NSNS 3-form flux is described by a sigma model on the $SL(2, R)$ group manifold \cite{Giveon:1998ns, deBoer:1998gyt, Teschner:1997fv, Teschner:1997ft, Teschner:1999ug, Maldacena:2000hw, Gukov:2004ym, Bars:1990rb, Brown:1986nw}. The string theory on the three dimensional AdS spacetime is related to a two dimensional conformal field theory residing on its one dimension lower boundary via the AdS/CFT correspondence. The central charge of the conformal field theory for large $k$ is given by the low energy gravity approximation $c \sim p k$. 

In the case in which the spatial direction of the boundary is compact, one can impose either periodic or anti-periodic boundary condition for the fermions of the boundary theory \cite{Coussaert:1993jp, Banados:1992gq, David:1999zb}. The periodic case gives the R sector.  The R sector ground state is described by massless BTZ which is a quotient of $AdS_3$. The anti-periodic case gives to the NS sector. The NS sector ground state is described by global $AdS_3$.

String theory on $AdS_3$ contains sectors of short and long strings \cite{Maldacena:2000hw, Dixon:1989cg}. The short strings are trapped in the interior of the AdS spacetime. They correspond to physical states that belong to the positive energy discrete representation of (the universal cover of) the $SL(2, R)$ current algebra and its spectral flow. Long strings can extend to the boundary and/or are near the boundary. They correspond to physical states that belong to the principal continuous representation of the $SL(2, R)$ current algebra and its spectral flow. 

There are evidences that support the idea that the boundary theory has the structure of a symmetric product \cite{Hosomichi:1999uj, Argurio:2000tb, Gaberdiel:2018rqv}. In particular the long string sector is believed to be described in the boundary theory by a conformal field theory with a symmetric product structure. The single-trace deformation gives further support to this idea in the long string sector. We will come back to this point shortly. However, in general, the boundary theory has not been fully understood, and in general it may not actually be described by a symmetric product theory.

The bosonic part of the world-sheet Lagrangian is 
\begin{equation}\label{eq:2.50}
S = {k\over 2\pi}\int d^2z (\partial\phi \bar\partial\phi + e^{2\phi}\partial\bar\gamma \bar\partial\gamma),
\end{equation}
where $(\phi, \gamma, \bar\gamma)$ are the coordinates on $AdS_3$. The boundary of $AdS_3$ is located at $\phi = \infty$. The action possesses an affine $SL(2, R)_L\times SL(2, R)_R$ symmetry, with independent generators for the left and right movers. The left mover world-sheet currents are given by
\begin{equation}\label{eq:2.51}
J^- = e^{2\phi}\partial\bar\gamma, \quad J^+ = -2\gamma\partial\phi - \partial\gamma + \gamma^2e^{2\phi}\partial\bar\gamma, \quad J^3 = \gamma e^{2\phi}\partial\bar\gamma -  \partial\phi,
\end{equation}
and similar expressions for the right movers. The left-moving currents $J^-, J^3$ and $J^+$ give in spacetime the left-moving global conformal charges $L_{-1}, L_0$ and $L_{+1}$, respectively, and similarly for the right-movers. This can be generalized to the supersymmetric case. In our discussions we only consider the bosonic part.

We introduce the auxiliary fields $\beta, \bar\beta$. The Lagrangian becomes, taking into account effects on the measure of the functional integral and rescaling the scalar fields $\phi, \gamma, \bar\gamma$,
\begin{equation}\label{eq:2.52}
S = {1\over 4\pi}\int d^2z (\partial\phi \bar\partial\phi + \beta\bar\partial\gamma + \bar\beta\partial\bar\gamma - \beta\bar\beta e^{-2\phi/\alpha_+} - {2\over \alpha_+}\phi\sqrt{g}R),
\end{equation}
where $\alpha_+ = \sqrt{2(k - 2)}$ and $R$ is the curvature of the world-sheet. Quantum corrections produce the linear dilaton term \cite{David:1988hj, Ishibashi:2000fn, Hosomichi:2000bm}. This notation is useful to study the physics (in particular long strings) near the boundary or large positive $\phi$ where the string coupling is small and the theory becomes free. The left mover currents in this notation are given by (see for example \cite{deBoer:1998gyt, Hofman:2004ny, Hosomichi:2000bm})
\begin{equation}\label{eq:2.53}
J^- = \beta, \quad J^+ = \beta\gamma^2 - 2\alpha_+ \gamma\partial\phi - \alpha^2_+ \partial\gamma, \quad J^3 = \beta\gamma - \alpha_+\partial\phi.
\end{equation}

At large $k$ a single long string is described by a conformal field theory ${\mathcal M}$ with central charge $c = 6k$ \cite{Seiberg:1999xz}. Since the number of NS1-branes $p$ is fixed, we have at most $p$ long strings. The CFT of the long string sector is believed to be given by the symmetric product \cite{Hosomichi:1999uj, Argurio:2000tb, Chakraborty:2019mdf}
\begin{equation}\label{eq:2.54}
S^p ({\mathcal M}) := {\mathcal M}^p/S_p,
\end{equation}
where ${\mathcal M}$ is the CFT of a single long string and $S_p$ denotes the symmetric group, that is, the permutation group of $p$ elements.  The central charge of $S^p({\mathcal M})$ is given by $c = 6pk$. A twisted sector of the theory $S^p( {\mathcal M})$ can describe fewer than $p$ long strings. The reason is that the twisted sectors are organized by the conjugacy classes of the symmetric or orbifold group $S_p$ \cite{Hosomichi:1999uj, Vafa:1994tf, Dijkgraaf:1996xw}. %This is so because

The background ${\mathcal M}_3$ (\ref{eq:2.48}) corresponds to adding to the world-sheet Lagrangian \ref{eq:2.52} a term proportional to the bilinear current $J^-{\bar J}^-$ \cite{2017arXiv170105576G}. The deformed action is
\begin{equation}\label{eq:2.55}
S_{\alpha_1} :=  {1\over 4\pi}\int d^2z \left(\partial\phi \bar\partial\phi + \beta\bar\partial\gamma + \bar\beta\partial\bar\gamma - \beta\bar\beta e^{-2\phi/\alpha_+} - {2\over \alpha_+}\phi\sqrt{g}R + \alpha_1 \beta\bar\beta\right),
\end{equation}
where $\alpha_1$ is the deformation coupling. The deformation breaks the $SL(2, R)_L \times SL(2, R)_R$ current algebra. At $\alpha_1 = 0$ the zero mode corresponding to $J^-$(${\bar J}^-$) gives rise to the spacetime Virasoro generator $L_{-1}$(${\bar L}_{-1}$), and thus, it has the spacetime scaling dimension $(1, 0)$($(0, 1)$). Therefore, the coupling $\alpha_1$ has spacetime scaling dimension $(-1, -1)$ similar to the $T{\bar T}$ deformation coupling. %S + \alpha_1 \int d^2z J^- {\bar J}^-  =

String theory in $AdS_3 \times {\mathcal M}$ contains an integrated vertex operator $D(x, {\bar x})$ that depends on the coordinates $(x, \bar x)$ of the boundary spacetime, and thus, the operator $D(x, {\bar x})$ lives in the boundary theory \cite{Kutasov:1999xu}. This operator is related to the deformation through \cite{2017arXiv170105576G}
\begin{equation}\label{eq:2.56}
\int d^2z J^-{\bar J}^- = t \int d^2xD(x, {\bar x}), 
\end{equation}
here $t$ is some constant. Therefore, deformation of the world-sheet action by adding the operator $\int d^2 z J^-{\bar J}^-$ is described in the dual boundary theory by adding to the corresponding action the operator $\int d^2x D(x, {\bar x})$. The operator $D(x, {\bar x})$ has spacetime scaling dimension $(2, 2)$. Thus, it is irrelevant. %We next discuss the operator in the boundary theory that gets identified with the integrated vertex operator $D(x, {\bar x})$.

In the symmetric product theory $S^p ({\mathcal M})$ that describes the long string sector, the operator $D(x, {\bar x})$ is identified with \cite{Chakraborty:2019mdf}\footnote{To be precise \begin{equation}\label{eq:2.57pp}
t D(x, {\bar x}) = c\lim_{p \to \infty}\sum_{i = 1}^{p}\int_0^td\alpha ({\rm T\bar T})^{(\alpha)}_i(x, {\bar x}),
\end{equation}
where $\alpha$ labels the trajectory and $c$ is a constant. Thus, from the single-trace deformation perspective, the $T{\bar T}$ deformation can be equivalently viewed as an RG upflow.}
\begin{equation}\label{eq:2.57}
D(x, {\bar x}) \equiv \sum_{i = 1}^{p} (T\bar T)_i(x, {\bar x}),
\end{equation}
where $(T{\bar T})_i$ is the $T{\bar T}$ operator in the $i$th copy of the CFT ${\mathcal M}$. Therefore, it is given by taking the sum hence the name single-trace $T{\bar T}$ operator. $D(x, {\bar x})$ transforms under $T(x)$ and ${\bar T}({\bar x})$ of $S^p ({\mathcal M})$ as a quasi-primary operator of dimension $(2, 2)$. Its OPE's with $T(x)$ and ${\bar T}({\bar x})$ of $S^p ({\mathcal M})$ is similar to that of $T{\bar T}(x, {\bar x})$ of $S^p ({\mathcal M})$ however $D(x, {\bar x})$ is a single-trace operator and thus, it is distinct. Note that the $T{\bar T}$ operator of the theory $S^p ({\mathcal M})$ is
\begin{equation}\label{eq:2.577z}
T{\bar T}(x, {\bar x}) = \left(\sum_{i = 1}^{p} T_i\right)\left( \sum_{j = 1}^p{\bar T}_j\right),
\end{equation}
where $T_i$ and ${\bar T}_i$ are the components of the energy momentum tensor in the $i$th copy of the CFT ${\cal M}$. Therefore, it is obtained by taking product of two sums or traces hence the name double-trace $T{\bar T}$ operator.

Therefore, it follows from the identification \ref{eq:2.57} that the spectrum for the untwisted sector of the symmetric product theory $S^p ({\mathcal M})$ \ref{eq:2.54} should be equal to that of the $T{\bar T}$ deformed spectrum of the copy ${\mathcal M}$. We next show this.

\subsection{Deformed spectrum}

We next show that string theory on the deformed background ${\mathcal M}_3$ \ref{eq:2.48} is equivalent to string theory on $AdS_3 $ with twisted boundary conditions for the coordinates or fields. We use this interpretation to compute the spectrum in the untwisted sector of the deformed symmetric product theory.

We compactify the spatial coordinate of the boundary. Therefore, the topology of the string background is $\mathbb{R}^{1, 1}\times S^1$. In the boundary theory the fermions can be either periodic or anti-periodic along the compact direction. This leads to two distinct sectors. The R sector is described by BTZ solutions and the NS sector is described by global AdS \cite{Coussaert:1993jp, Banados:1992gq, David:1999zb}.

The bosonic part of the deformed world-sheet action is described by the action
\begin{equation}\label{eq:2.58}
S = \frac{1}{2\pi}\int d^2 z\left(\partial \phi\bar\partial \phi + \beta\bar\partial\gamma + \bar{\beta}\partial\bar\gamma  + \left(\alpha_1 - e^{-2\phi/\sqrt{k}}\right)\beta\bar\beta \right),
\end{equation}
where
\begin{equation}\label{eq:2.59}
z :=  \tau - \sigma, \quad {\bar z} :=  \tau + \sigma,
\end{equation}
$\sigma$ is the spatial coordinate along the string and $\tau$ describes its propagation in time. Note that we can set the deformation coupling $\alpha_1 = 1$ by shifting the radial coordinate $\phi$. The boundary is at $\phi = +\infty$. The null coordinates $\gamma$ and $\bar\gamma$ are given by
\begin{equation}\label{eq:2.60}
\gamma = x - t, \quad \bar\gamma = x + t, \quad x \sim x + 2\pi.
\end{equation}

The deformation \ref{eq:2.58} can also be understood as a T-duality-shift-T-duality (TsT) (or more generally ${\cal O}(d, d)$) transformation \cite{Lunin:2005jy, Giveon:1994fu, Forste:1994wp, Araujo:2018rho, Sfondrini:2019smd}. TsT transformation is a systematic procedure to generate a new supergravity solution. Such TsT transformations can be equivalently viewed as twisted boundary conditions \cite{Lunin:2005jy, Sfondrini:2019smd, vanTongeren:2018vpb, Alday:2005ww, Frolov:2005dj, Azeyanagi:2012zd}. In this view the background spacetime is $AdS_3$. In what follows we derive the twisted boundary conditions corresponding to \ref{eq:2.58}. We follow a similar approach used in \cite{Azeyanagi:2012zd}. We then use the twisted boundary conditions to determine the spectrum in the deformed boundary theory.

We vary the action to get the equations of motion for the bosonic string. We find
\begin{equation}\label{eq:2.61}
\partial\bar\partial\phi -e^{-2\phi/\sqrt{k}}\beta\bar\beta = 0,
\end{equation}
\begin{equation}\label{eq:2.62}
\bar\partial\gamma + (\alpha_1 - e^{-2\phi/\sqrt{k}})\bar\beta = 0,
\end{equation}
\begin{equation}\label{eq:2.63}
\partial\bar\gamma + (\alpha_1 - e^{-2\phi/\sqrt{k}})\beta  = 0,
\end{equation}
\begin{equation}\label{eq:2.64}
\bar\partial\beta = 0 ,
\end{equation}
\begin{equation}\label{eq:2.65}
\partial\bar\beta = 0.
\end{equation}
The holomorphic (left-moving) and anti-holomorphic (right-moving) energy momentum tensor components are given at classical level by 
\begin{equation}\label{eq:2.66}
-T = (\partial\phi)^2  + \beta\partial\gamma , \qquad -{\bar T} = (\bar\partial\phi)^2  + \bar\beta \bar\partial\bar\gamma.
\end{equation}
At quantum level normal ordering generates an improvement term for $\phi$.

We now make the following redefinition of the fields
\begin{equation}\label{eq:2.67}
  \bar\partial\hat\gamma = \bar\partial \gamma + \alpha_1\bar\beta , \quad \partial\hat{\bar\gamma}= \partial\bar\gamma + \alpha_1\beta,  \quad \partial\hat\gamma = \partial \gamma ,  \quad {\bar\partial}\hat{\bar\gamma }= {\bar\partial }{\bar \gamma }.
\end{equation}
The solutions to these equations are given by
\begin{equation}\label{eq:2.68}
\hat\gamma =  \gamma + \alpha_1\mu_+ , \quad \hat{\bar\gamma} = \bar\gamma + \alpha_1\mu_- , 
\end{equation}
where 
\begin{equation}\label{eq:2.69}
\beta = \partial\mu_-, \quad \bar\beta = \bar\partial\mu_+, 
\end{equation}
are the bosonizations of the currents $J^- := i\beta$ and ${\bar J}^- := -i\bar\beta$.

In terms of the new hatted variables we note that the local world-sheet dynamics becomes

\begin{equation}\label{eq:2.70}
\partial\bar\partial\phi -e^{-2\phi/\sqrt{k}}\beta\bar\beta = 0,
\end{equation}
\begin{equation}\label{eq:2.71}
\bar\partial\hat\gamma - e^{-2\phi/\sqrt{k}}\bar\beta = 0,
\end{equation}
\begin{equation}\label{eq:2.72}
\partial\hat{\bar\gamma} - e^{-2\phi/\sqrt{k}}\beta  = 0,
\end{equation}
\begin{equation}\label{eq:2.73}
\bar\partial\beta = 0,\qquad \partial\bar\beta = 0.
\end{equation}
 We also note that the components of the energy momentum tensor take the following forms
\begin{equation}\label{eq:2.74}
-T = (\partial\phi)^2  + \beta\partial\hat\gamma , \qquad -{\bar T} = (\bar\partial\phi)^2  + \bar\beta \bar\partial\hat{\bar\gamma}. 
\end{equation}
The improvement term is not affected since it only depends on $\phi$. Therefore, the local dynamics of the string in the deformed background is given by the $SL(2, R)$ WZW sigma model at level $k$. This is a particularly important observation since we can now make use of the $SL(2, R)_L \times SL(2, R)_R$ affine current algebra to construct vertex operators, obtain spectrum and compute correlation functions.%full

We define the zero mode momenta $p_L$ and $p_R$ as 
\begin{equation}\label{eq:2.75}
 p_L := -\frac{1}{2\pi}\oint d\sigma \partial\mu_-,   \quad p_R := -\frac{1}{2\pi}\oint d\sigma\bar\partial\mu_+,
\end{equation}
where the contour integral is at a fixed world-sheet time. The momenta $p_R$ and $p_L$ are related to the spacetime (boundary) energy $\bar E$ and momentum $\bar P$ via the relations (see, for example, \cite{Troost:2002wk})
\begin{equation}\label{eq:2.76}
 {\bar P} = p_L + p_R, \quad {\bar E} = p_R - p_L . 
\end{equation}
In the case in which $x$ is not compact the momentum $\bar P$ is not quantized.

We note that, at fixed world-sheet time, using the identity 
\begin{equation}\label{eq:2.77}
\oint dz\partial\xi + \oint d\bar z \bar\partial \xi =  \oint d\sigma \partial_{\sigma }\xi = \xi(\sigma + 2\pi) - \xi(\sigma) :=\Delta \xi, 
\end{equation}
we find the following twisted boundary conditions 
\begin{equation}\label{eq:2.78}
 \Delta \hat\gamma = \Delta \gamma  - 2\pi\alpha_1 p_R := 2\pi{\hat w}_R(p_R), \quad \Delta \gamma = 2\pi w_R, 
\end{equation}
\begin{equation}\label{eq:2.79}
\Delta\hat{\bar\gamma} = \Delta\bar\gamma  + 2 \pi\alpha_1 p_L := 2\pi{\hat w}_L(p_L), \quad \Delta \bar\gamma = 2\pi w_L, 
\end{equation}
where $w_L $ ($ = w_R := w$) is the winding number of the string which counts the number of times the string winds around the compact spatial direction. The winding number $w$ can be positive or negative depending on the string orientation. 

Therefore, string theory on the deformed background ${\mathcal M}_3$ (\ref{eq:2.48}) is the same as string theory on the undeformed background $AdS_3$ but with twisted boundary conditions \ref{eq:2.78} and \ref{eq:2.79}. As we will show shortly, this is equivalent to a momentum dependent spectral flow. We also note that in general ${\hat w}_L \neq {\hat w}_R$. Thus, the fields ${\hat\gamma}$ and $\hat{\bar\gamma}$ in general describe non-local dynamics. In chapter three and four we examine the non-local behavior of the theory in more detail using correlation functions and entanglement entropy.

We are in particular interested in the long string sector spectrum and therefore we use the free field representation. In the free field approximation we have 
\begin{equation}\label{eq:2.80}
\phi(z, \bar z)\phi(w, \bar w) \sim -\frac{1}{2}\ln |z - w|^2, 
\end{equation}
\begin{equation}\label{eq:2.81}
\partial\mu_-(z)\hat \gamma(w) \sim -\frac{1}{z - w}, \qquad \partial\hat\gamma(z)\mu_-(w) \sim -\frac{1}{z - w},
\end{equation}
and similar OPEs for the fields $\mu_+, \hat{\bar\gamma}$.

We now construct world-sheet vertex operators $\hat {\cal V}^{(w)}_{p_L, p_R}$ that create states in a given winding sector $w$ with definite momenta $p_L$ and $p_R$. The monodromy conditions \ref{eq:2.78} and \ref{eq:2.79} imply the OPEs
\begin{equation}\label{eq:2.82}
\hat\gamma (z_1, \bar z_1) \hat{\cal V}^{(w)}_{p_L, p_R}(z_2, \bar z_2) \sim  i\left(-w +  \alpha_1 p_R\right)\ln(z_1 - z_2) \hat{\cal V}^{(w)}_{p_L, p_R}(z_2, \bar z_2),
\end{equation}
\begin{equation}\label{eq:2.83}
\hat{\bar\gamma} (z_1, \bar z_1) \hat {\cal V}^{(w)}_{p_L, p_R}(z_2, \bar z_2) \sim -i\left(w + \alpha_1 p_L\right)\ln(\bar z_1 - \bar z_2) \hat{\cal V}^{(w)}_{p_L, p_R}(z_2, \bar z_2),
\end{equation}
%which follow from the twisted boundary conditions \ref{eq:2.82} and \ref{eq:2.83} are satisfied. This requirement corresponds the dressing
This corresponds the dressing
\begin{equation}\label{eq:2.84}
 \hat{\cal V}^{(w)}_{p_L, p_R} =   \hat V^{(w)}_{p_L, p_R}   e^{-i \alpha_1 p_R\mu_-} e^{i\alpha_1p_L\mu_+},
\end{equation}
where $\hat V^{(w)}_{p_L, p_R} $ is the world-sheet vertex operator on the undeformed background $AdS_3$ \cite{Parsons:2009si, Troost:2002wk}. The vertex operator $\hat V^{(0)}_{p_L, p_R} := \hat V_{p_L, p_R}$ is obtained by taking the Fourier transform of the primary operator $\hat V_{\hat h, \hat h}$, see appendix \ref{app_one}. %Hofman:2004ny, Hikida:2000ry, Evans:1998qu, Giveon:2017myj,
It has, in the large $\phi$ and with our definitions in \ref{eq:2.75}, the form
\begin{equation}\label{eq:vx}
\hat V^{(w)}_{p_L, p_R} = e^{ip_L\hat\gamma +iw \mu_-} e^{-ip_R\hat{\bar\gamma} +iw\mu_+}  f_{\hat h}(\phi),
\end{equation}
where $f_{\hat h}(\phi) \sim e^{2{(\hat h} - 1)\phi/\sqrt{k}}$ and the constant $\hat h$\footnote{We take the left $\hat h$ and right $\hat{\bar h}$ to be equal.} labels the (continuous) representations of the current algebra. Note that the deformed vertex operators $\hat{\cal V}^{(w)}_{p_L, p_R}$ are mutually local since $\bar P$ is quantized and thus have no branch cut. %is the spacetime scaling dimension of $\hat V^{(0)}_{p_L, p_R} := \hat V_{p_L, p_R}$

The conditions \ref{eq:2.82} and \ref{eq:2.83} (or equivalently the dressing \ref{eq:2.84} with \ref{eq:vx}) are equivalent to the spectral flow
\begin{equation}\label{eq:2.88}
{\cal L}_m =   L_m - \left( \alpha_1 p_R - w\right)J^-_m, 
\end{equation}
\begin{equation}\label{eq:2.89}
\bar{\cal L}_m =  {\bar L}_m - \left(\alpha_1p_L + w\right)\bar{J}^-_m, 
\end{equation}
where $ L_m$ (${\bar L}_m$) are the modes of the left (right) moving component of the undeformed world-sheet energy momentum tensor and $J^{-}_m$ (${\bar J}^{-}_m$) are the modes of the left (right) moving (undeformed) world-sheet current $J^-$(${\bar J}^-$).

The Virasoro constraint ${\cal L}_0 = 1$ ($\bar{ \cal L}_0 = 1$) and the relations in \ref{eq:2.76} lead to the equation
\begin{equation}\label{eq:2.90}
{\alpha_1\over 4} \left({\bar P}^2 - {\bar E}^2\right) + {w\over 2}\left({\bar E} - {\bar E}_0\right) = 0, \quad {\bar E}_0 := {\bar E}(\alpha_1 = 0), \quad \partial_{\alpha_1}{\bar P} = 0.
\end{equation}
This gives the spacetime spectrum
 \begin{equation}\label{eq:2.91}
 {\bar E}(\lambda) = {\bar P} -{1\over 2A}\left(B + \sqrt{B^2 - 4AC}\right),
 \end{equation}
where
\begin{equation}\label{eq:2.92}
A = -{1\over 4}\lambda, \quad B = -w - {1\over 2}\lambda {\bar P} , \quad C = w\left({\bar E}_0 - {\bar P}\right), %= 2\left( h_w - \frac{wc}{24}\right)
\end{equation}
and
\begin{equation}\label{eq:2.93}
\alpha_1 = -{1\over 2}\lambda, \quad \lambda =  {8\pi t\over R^2}, \quad {\bar E} = {R{\mathcal E}\over 2\pi}, \quad {\bar P} = {R P\over 2\pi }.
\end{equation}

We note that when $w = 1$ the single-trace $T{\bar T}$ deformed spectrum with the relations \ref{eq:2.93} matches the spectrum of a $T{\bar T}$-deformed CFT obtained earlier in \ref{eq:2.34}. Recall that the CFT $\mathcal M$ in \ref{eq:2.54} is the dual theory that describes a single long string. Therefore, the spectrum of (single particle) states in the untwisted or $w = 1$ sector of the dual single-trace deformed symmetric product is given by the spectrum of the  $T{\bar T}$ deformed theory ${\mathcal M}$ in \ref{eq:2.54}. Thus, the single-trace deformation supports the identification \ref{eq:2.57}. This analysis can be extended for theories with currents \cite{Apolo:2018qpq}. See also \cite{Giveon:2017myj, Giveon:2019fgr}. 

In the case in which the spatial direction is not compact the dressed un-flowed vertex operator $ \hat{\cal V}_{p_L, p_R}$ (\ref{eq:2.84}) has the world-sheet scaling dimension given by
\begin{equation}\label{eq:2.85}
 -\frac{  \hat h( \hat h - 1)}{k}  - \alpha_1 p^2, 
\end{equation}
where $p^2:= p_L p_R$.

\chapter{Correlation functions}

This chapter is based on the paper \cite{Asrat:2017tzd}.

In this chapter the question that we address is concerning how the non-locality and UV behavior of the deformed boundary theory is manifested on the analytic structure of correlation functions of certain class of (gauge invariant) operators \cite{Aharony:2007rq}. We in particular compute two point function in the case where the spatial direction is not compact. We use the momentum dependent spectral flow. 

The class of operators we consider has in the IR undeformed CFT the form
\begin{equation}\label{eq:2.97x}
{\mathcal O}_{h, h}(x) = \int d^2z V_{h, h}(x; z){\mathcal U}(z),
\end{equation}
where $V_{h, h}$ (see \ref{eq:2.85z1}) is a vertex operator in $AdS_3$ and ${\mathcal U}(z)$ is a vertex operator in the internal CFT on the compact manifold ${\mathcal N}$. The coordinate $x$ labels the boundary and the coordinate $z$ labels the world-sheet or the sphere. The operator $V_{h, h}$ has the spacetime scaling dimension $(h, h)$. The world-sheet scaling dimension $(\Delta_{\mathcal U}, \Delta_{\mathcal U})$ of the operator ${\mathcal U}$  is related to the world-sheet scaling dimension $(\Delta_h, \Delta_h)$ of the operator $V_{h, h}$ through the mass-shell condition
\begin{equation}\label{eq:2.98}
\Delta_h + \Delta_{\mathcal U} = 1, \quad \Delta_h = -{h(h - 1)\over k}.
\end{equation}

In $AdS_3$ (tree-level or sphere) two point function of vertex operators $V_{h, h}$ in position space in a particular normalization is given by \cite{Teschner:1999ug, Maldacena:2001km},
\begin{equation}\label{eq:2.87}
\langle   V_{h, h}(x, z) V_{h', h'}(y, w)\rangle = \delta(h -  h'){B(h)\over |x - y|^{4 h}|z - w|^{4 \Delta_h}},
\end{equation}
where the function $B$ is
 \begin{equation}
B(a) = {k\over\pi}\nu^{1 - 2a}\gamma\left(1- \frac{2a - 1}{k}\right),  \quad \gamma(a) = \frac{\Gamma(a)}{\Gamma(1 - a)},
\end{equation} 
$\nu$ is some known function of the level $k$. In general, $B(h)$ can be made smooth by choosing a particular normalization for the operator $V_{h, h}$ (of course, unless it has a natural normalization), and we shall assume that this is indeed the case. We normalize ${\mathcal U}$ such that $\langle {\mathcal U}(z){\mathcal U}(w) \rangle = 1/|z - w|^{\Delta_{\mathcal U}}$. Therefore, the physics is captured by the simple power functions with the exponents $4h$ and $4\Delta_h$. The two point function of the operators ${\mathcal O}_{h, h} := {\mathcal O}$ \ref{eq:2.97x} is then given by
\begin{equation}\label{eq:2.87777}
\langle   {\mathcal O}(x) {\mathcal O}(y)\rangle = { \delta(0)\over V}{B(h)\over |x - y|^{4 h}} =  {D(h)\over |x - y|^{4 h}},
\end{equation}
where $V$ is the volume of the conformal group on the sphere and $D(h) = (2h - 1)B(h)$.

In the deformed theory the operator ${\mathcal O}$ deforms as %according to the momentum dependent spectral flow \ref{eq:2.84} as
\begin{equation}\label{eq:2.97}
{\hat {\mathcal O}}_{\hat h, \hat h}(x) = \int d^2z {\hat {\cal V}}_{\hat h, \hat h}(x; z){\mathcal U}(z),
\end{equation}
where ${\hat {\cal V}}_{\hat h, \hat h}$ is the deformed vertex operator. The deformation does not break the world-sheet conformal symmetry and in the momentum dependent spectral flow interpretation we saw that a vertex operator in the deformed theory has the world-sheet conformal dimension \ref{eq:2.85}. The conformal dimension does not depend on what interpretation one chooses. Thus, in the deformed theory the mass-shell condition takes the form
\begin{equation}\label{eq:2.97X} 
\hat{\Delta}_{{\hat h}} +  \Delta_{\mathcal U}  = 1, \quad \hat{\Delta}_{{\hat h}} = -\frac{  \hat h( \hat h - 1)}{k} + {1\over 2}\lambda p^2.
\end{equation}
This equation can be solved for $\hat h := \hat h_{p^2}(\lambda)$ as a function of the deformation coupling $\lambda$. This gives using \ref{eq:2.98} the flow equation
\begin{equation}\label{eq:2.86} 
2 \hat h_{p^2}(\lambda) - 1 = \sqrt{(2h- 1)^2 + 2k\lambda p^2},  \quad \hat h_{p^2}(\lambda = 0) := h.
\end{equation}
%The spacetime scaling dimension of the deformed operator $\hat{\mathcal O}$ is given by ${\hat h}_{p^2}$ \ref{eq:2.86}.  
As we also saw earlier, the local dynamics of strings in the deformed background is described by the $SL(2, R)$ WZW sigma model at level $k$. Therefore, (the un-flowed $w = 0$) two point function of vertex operators in the deformed theory is also given by \ref{eq:2.87} with $\Delta_h$ replaced by $\hat{\Delta}_{{\hat h}}$ and $h$ replaced by ${\hat h}_{p^2}$. The exponent $\hat h_{p^2}$ is related with the momentum $p^2$ via the relation $\ref{eq:2.86}$\footnote{Note that this implies a different choice of normalization introduces a different momentum or equivalently coupling dependence.}. Thus, in the next section, we Fourier transform the  two point function \ref{eq:2.87777} with $h$ replaced by ${\hat h}_{p^2}$ and transform it back to position space so that it only depends on position coordinates. We consider the two point function separately in momentum and position spaces. See \cite{Asrat:2017tzd} for detailed discussions where coset construction is used to study correlation functions. 

\section{Two point function in momentum space}

We will do the calculation in Euclidean space. In momentum space we find using \ref{eq:2.87777} for the two point function 
\begin{equation}\label{eq:2.101}
\langle {\hat {\mathcal O}}(p){\hat {\mathcal O}}(-p)\rangle = \pi D({\hat h}_{p^2})\gamma(1 - 2{\hat h}_{p^2})\left(p^2\over 4\right)^{2{\hat h}_{p^2} - 1},
\end{equation}
where $D(x) = (2x - 1)B(x)$ and it is a smooth function. In particular for $\lambda = 0$ this gives the undeformed two point function in momentum space  
\begin{equation}\label{eq:2.102}
\langle { \mathcal O}(p){ \mathcal O}(-p)\rangle = \pi D( h)\gamma(1 - 2 h)\left(p^2\over 4\right)^{2h - 1}.
\end{equation}
The momentum space two point function \ref{eq:2.102} as a function of $h$ has poles for $2h - 1 = n \in \mathbb{Z}_+$. They appear in the multiplicative factor that does not depend on the momentum, and they can be removed or cancelled by appropriate renormalization of the vertex operators. In this case where $2h - 1 = n \in {\mathbb Z}_+$ the high momentum limit of the regularized two point function behaves $p^{2n}\log\left(p^2/ \mu^2\right)$ where $\mu$ is some mass scale (see for example \cite{Bzowski:2013sza, Gurarie:1993xq, Hogervorst:2016itc, Petkou:1999fv}). In all other cases the two point function behaves as power law $\left(p^2\right)^{2h - 1}$. It follows, therefore, that the non-trivial analytic properties of the two point function \ref{eq:2.101} are fully captured by the function ${\hat h}_{p^2}$ \ref{eq:2.86}.

In Euclidean momentum space the two point function \ref{eq:2.101} is real and has no branch cut. In the (strictly) high Euclidean momentum limit it exhibits the hyper-power law growth $p^{\alpha p}$ for some known positive constant $\alpha$. Such growth is very different from the power law behavior we saw earlier. However, in Minkowski momentum space the two point function is real and has no branch cut only for spacelike momenta. For timelike momenta the flow equation \ref{eq:2.86} (or equivalently the momentum two point function \ref{eq:2.101}) has a branch cut with branch points given by
\begin{equation}\label{eq:2.103}
p_0^2 := -{(2h - 1)^2\over 2k\lambda}. 
\end{equation}.

In general branch cut in correlation functions or OPEs signals non-locality and/or ambiguity. We next consider the two point function in position space.

\section{Two point function in position space}

In position or real space the two point function is given by
\begin{equation}\label{eq:2.104}
\langle \hat{\mathcal O}(y)\hat {\mathcal O}(0) \rangle = 2\pi \int_0^{\infty} d\mu^2 \rho(\mu^2; \lambda) K_0(\mu|y|),
\end{equation}
where $K_0$ is the Fourier transform of the propagator
\begin{equation}\label{eq:2.105}
\int d^2p {e^{i p\cdot y}\over p^2 + \mu^2} = 2\pi K_0(\mu  |y|),
\end{equation}
and $\rho(\mu^2; \lambda)$ is the spectral density. Note that $K_0$ is equal to the modified Bessel function. In the regime $\mu |y| \gg 1$ the Bessel function decays exponentially $K_0(\mu |y|) \sim e^{-\mu |y|}/\sqrt{\mu |y|}$, and in the regime $\mu |y| \ll 1$ it diverges logarithmically $K_0 \sim -\rm{log} (\mu |y|)$. The spectral density is defined in Kallen-Lehmann representation by the equation 
\begin{equation}\label{eq:2.106}
\langle {\hat {\mathcal O}}(p){\hat {\mathcal O}}(-p)\rangle = \int_0^{\infty} d\mu^2 {\hat\rho(\mu^2; \lambda)\over \mu^2 + p^2}.
\end{equation}
That is, the two point function is obtained by summing the free scalar momentum space propagator or two point function over all masses $\mu$ with a weight or probability density given by the spectral density $\hat\rho(\mu^2; \lambda)$\footnote{This identification of $\mu$ with mass should not be taken literally; it is a mere observation. For example, in a CFT there are no mass scales.}. Using the momentum space two point function \ref{eq:2.101} in the definition \ref{eq:2.106} we find that the spectral density $\hat\rho$ is given by
\begin{equation}\label{eq:2.107}
\hat\rho(\mu^2; \lambda) = \pi {D(\hat{h}_{-\mu^2})\over \Gamma^2(2{\hat h}_{-\mu^2})}\left({\mu^2\over 4}\right)^{2{\hat h}_{-\mu^2} - 1}.
\end{equation}
For a CFT or at $\lambda = 0$ this gives
\begin{equation}\label{eq:2.107y}
\rho(\mu^2) = \pi {D(h)\over \Gamma^2(2h)}\left({\mu^2\over 4}\right)^{2h - 1}.
\end{equation}
We note that spectral density is positive and continuous. It also does not contain a delta function at $\mu = 0$\footnote{See \cite{Duetsch:2002hc} for a discussion in relation to Wightman and generalized free fields.}.
 
The spectral density \ref{eq:2.107} is complex for $\mu >  \mu_0$ where $\mu_0^2 = -p_0^2$ is given by \ref{eq:2.103}. It has a real and an imaginary part. The interpretation of the imaginary part of the density is not completely clear, however, we expect to be related to the non-local behavior of the theory. It follows, therefore, that the two point function \ref{eq:2.104} in position space is complex. In perturbation theory, keeping only the leading logarithmic term at each order in the deformation coupling, the two point function \ref{eq:2.104} in position space is given in compact form as
\begin{equation}\label{eq:2.108}
\langle \hat{\mathcal O}(y)\hat {\mathcal O}(0) \rangle  =  e^{\frac{4\lambda k}{2h - 1} \ln  \left(m^2|y|^{2}\right) \partial_y\partial_{\bar{y}}}\langle {\mathcal O}(y) {\mathcal O}(0) \rangle,
 \end{equation}
 where $m$ is some arbitrary normalization scale and the two point function in the undeformed theory is
\begin{equation}\label{eq:2.108x}
\langle {\mathcal O}(y) {\mathcal O}(0) \rangle  =  \left(2\pi\right)^2 |y|^{-4h},
 \end{equation} 
 in some normalization. We in particular note that in perturbation theory the two point function is real. In the regime where $\mu_0|y| \gg 1$ or equivalently small $\lambda/|y|^2$ we find that the imaginary part goes like $e^{-\sqrt{|y|^2/ \lambda}}$. Thus, the imaginary part of the two point function is non-perturbative in the deformation coupling $\lambda/|y|^2$. A simple way to see this is by considering the large order behavior of the two point function (see \cite{LeGuillou:1990nq}). We write the two point function \ref{eq:2.104} as
 \begin{equation}\label{eq:2.108xx}
\langle \hat{\mathcal O}(y)\hat {\mathcal O}(0) \rangle = {1\over |y|^{4h}}\sum_{n = 0}^\infty c_n \left({\lambda\over |y|^2}\right)^n.
 \end{equation}
The large order behavior of the coefficient $c_n$ is related to the imaginary part. One obtains
\begin{equation}\label{eq:2.108xxx}
c_n \sim n^\alpha \cdot e^{\beta n} \cdot (2n)!,
 \end{equation}
where $\alpha$ and $\beta$ are some constants. Such double factorial growth is common in string theory and it corresponds to weak coupling non-perturbative effects \cite{Shenker:1990uf}. The imaginary part is directly related to the imaginary part of the spectral density, and thus, it is related to the non-local behavior of the theory.

\chapter{Entanglement entropy and Entropic \texorpdfstring{$c$}{c}-function}

In this chapter we study the holographic entanglement entropy of an interval in a QFT obtained by deforming a holographic two--dimensional CFT via a general linear combination of single-trace irrelevant operators $T{\bar T}, \ J{\bar T}$ and $T{\bar J}$, and compute the Casin--Huerta entropic $c$--function. In the UV, for a particular combination of the deformation parameters, we find that the leading order dependence of the entanglement entropy on the length of the interval is given by a square root but not logarithmic term. Such power law dependence of the entanglement entropy on the interval length is quite peculiar and interesting. We also find that the entropic $c$--function is UV regulator independent, and along the RG upflow towards the UV, it is non--decreasing. We show that in the UV the entropic $c$--function exhibits a power law divergence as the interval length approaches a minimum finite value determined in terms of the deformation parameters. This value sets the non--locality scale of the theory. This chapter is based on the paper \cite{Asrat:2019end}.

\section{Single trace \texorpdfstring{$T{\bar T}$, $J{\bar T}$}{TTbar, JTbar} and \texorpdfstring{$T{\bar J}$}{TJbar} deformations}

String theory on the background $AdS_3 \times {\mathcal X}$ with Neveu--Schwartz two--form $B$ field contains a sector of long strings that extend to the boundary \cite{Maldacena:2000hw, Seiberg:1999xz, Argurio:2000tb, Maldacena:1998uz}. The effective theory of $N$ coincident long strings on $AdS_3 \times {\mathcal X}$ in the weak coupling regime is believed to be well described by the symmetric product theory \cite{ Argurio:2000tb, Chakraborty:2019mdf}
\begin{equation}\label{eq:2.109}
{\mathcal M}^N/S_N,
\end{equation}
where the conformal field theory ${\mathcal M}$ is the dual theory of a single long string \cite{Seiberg:1999xz}. Note that the symmetric product theory \ref{eq:2.109} is not equal to the full dual boundary (spacetime) theory since ${\mathcal M}$ only describes the long string sector. 

In \cite{Chakraborty:2019mdf} the authors considered deforming the Lagrangian of the world-sheet string theory on the background $AdS_3 \times {\mathcal X}$ by a general linear combination of truly marginal current--current operators 
\begin{equation}\label{eq:3.2}
\lambda J^- {\bar J}^- + \epsilon_+ K {\bar J}^- + \epsilon_- {\bar K} J^-.
\end{equation}
They showed that, in the long string sector, this deformation is equivalent to the deformation of the theory ${\cal M}$ (with spectral flow or winding number $w = 1$) by a general linear combination of the recently much studied irrelevant operators \cite{2004hep.th....1146Z, 2017NuPhB.915..363S, 2016JHEP...10..112C, 2018ScPP....5...48G}
\begin{equation}\label{eq:3.3}
-tT{\bar T} - \mu_+ J{\bar T} - \mu_ -{\bar J}T,
\end{equation}
where $J$ and ${\bar J}$ are the left and right moving  $U(1)$ currents, respectively, and $T$ and ${\bar T}$ are the left and right moving stress tensor components, respectively, of the conformal field theory ${\cal M}$. 

String theory on the background $AdS_3 \times {\cal X}$ contains a class of (integrated) vertex operators ${\bar A}, \ A$ and $D$ constructed in \cite{Kutasov:1999xu}, which depends on the coordinates $(x, {\bar x})$ of the dual conformal field theory. It is shown in \cite{2017arXiv170105576G, Chakraborty:2019mdf, Apolo:2018qpq} that the deformation \ref{eq:3.2} in the action amounts to adding a linear combination of the operators $\int d^2 x{\bar A}, \ \int d^2 x A$ and $\int d^2 x D$. In the boundary (spacetime) theory the operators ${\bar A}, \ A$ and $D$ have the same scaling dimensions as $T{\bar J}, \ J{\bar T}$ and $T{\bar T}$, respectively, however, they are single-trace operators, in the sense that, in the long string sector, they are interpreted as sum over the $N$ copies of the operators $T{\bar J}, \ J{\bar T}$ and $T{\bar T}$, respectively, of the field theory ${\cal M}$. 

The operators $J{\bar T}$ and $T{\bar J}$ have left and right scaling dimensions $(1, 2)$ and $(2, 1)$, respectively, and therefore, the deformation \ref{eq:3.3} results in a theory that breaks Lorentz invariance \cite{2018ScPP....5...48G, Chakraborty:2018vja, Apolo:2018qpq}.

The spacetime couplings $t, \ \mu_+$ and $\mu_-$ are related to the world-sheet dimensionless couplings $\lambda,\ \epsilon_+$ and $\epsilon_-$ via the relations \cite{Chakraborty:2019mdf},
\begin{equation}\label{eq:3.4}
t = \pi \alpha' \lambda, \quad \mu_{\pm} = 2\sqrt{2\alpha'}\epsilon_{\pm}, \quad \alpha' = l_s^2,
\end{equation}
where $\alpha'$ is the Regge slope, and $l_s$ is the intrinsic string length. 

The deformation \ref{eq:3.3} is irrelevant and therefore the couplings grow as we ascend the renormalization group. In general, under an irrelevant deformation of a quantum field theory, in the ultraviolet, the associated coupling is large and the description of the theory in terms of the original infrared degrees of freedom often breaks down. The theory also suffers from ambiguities and/or arbitrariness. It is also often the case that quantum corrections generate an infinite number of irrelevant operators. Under the deformation \ref{eq:3.3} it is shown, however, that the theory is solvable, in the sense that the spectrum on an infinite cylinder \cite{2004hep.th....1146Z, 2017NuPhB.915..363S, 2016JHEP...10..112C, 2018ScPP....5...48G, Giveon:2017myj, Giveon:2019fgr} and the partition function on a torus \cite{2018JHEP...10..186C, Dubovsky:2018bmo, Hashimoto:2019wct, Datta:2018thy, Aharony:2018bad, Aharony:2018ics} can be computed exactly. The theory also does not acquire new couplings. On a torus modular invariance can be employed to constrain the theory \cite{Asrat:2020jsh, Datta:2018thy, Aharony:2018bad, Aharony:2018ics}. In the case in which the coupling $t$ is positive and $\mu_\pm = 0$ the theory involves no ambiguities \cite{Aharony:2018bad, Aharony:2018ics}.  We  obtain in this chapter the exact entanglement entropy and entropic c--function in the deformation \ref{eq:3.3}. This increases the number of non--trivial quantities that one can exactly solve and study in this class of theories.

It is shown in \cite{Chakraborty:2019mdf} that in the space of couplings in which the combination
\begin{equation}\label{eq:3.5}
F =  {t\over \pi} - {(\mu_+ + \mu_-)^2\over 8},
\end{equation}
is positive, $F > 0$, the energies of states are real, and the density of states asymptotically exhibits Hagedorn growth. In the limit $F \to 0^+$ however the theory appears to be distinct. The density of states (in a fixed charge sector) asymptotically exhibits an intermediate growth between Cardy and Hagedorn growths. We also show later in the chapter that in this limit the Von Neumann entanglement entropy at short distances exhibits a square root area law correction but not  logarithmic correction. Such square root correction term is quite peculiar and interesting. In the case in which $F < 0$ the energies become complex above a scale fixed by the couplings and the corresponding bulk geometry is singular and/or it has either closed timelike curves or no timlike direction. The signature of the bulk metric switches signs beyond a finite radial distance where the singularity occurs. We will not consider this case in this chapter as it is not clear how to consistently apply the Ryu and Takayanagi holographic prescription and its covariant generalization. We comment on this later in the discussion section.

We now briefly mention the holographic proposals for the closely related double-trace deformations. In this class of deformations $T$ and $J$ are the left moving energy momentum tensor and current, respectively, of the full boundary theory. The right moving energy momentum tensor and current are denoted by $\bar T$ and $\bar J$ respectively. In these holographic proposals the coupling $t$ is negative and therefore $F < 0$. For negative $t$ with $\mu_\pm = 0$ the dual bulk spacetime of a $T{\bar T}$ deformed two dimensional holographic conformal field theory is proposed to be $AdS_3$ with a Dirichlet boundary at finite radial distance fixed by the coupling $t$ \cite{McGough:2016lol}. For either sign of $\mu_+$ (and $\mu_- = 0$ or vice versa) and $t = 0$ it is shown in \cite{Bzowski:2018pcy} that the dual bulk spacetime of a $J{\bar T}$ (or $T{\bar J}$) deformed two dimensional holographic conformal field theory is $AdS_3$ with boundary conditions that mix the metric and a gauge field dual to the current $J$ (or $\bar J$). In either of these cases in the field theory side we have states with complex energies.

In this chapter we study the (Von Neumann) entanglement entropy for a spatial interval of length $L$ in the deformed (full) spacetime theory with $F \geq 0$ from its bulk string theory description and compute the Casin--Huerta entropic $c$--function. We study the monotonicity property of the entropic c--function along the renormalization group upflow and its independence of regularization scheme that one introduces to regularize the ultraviolet divergence of entanglement entropy. This provides further support that the renormalization group upflow is under better control.  It also gives important insight into the nature of the theory in the ultraviolet as it is not governed by an ultraviolet fixed point \cite{Dubovsky:2017cnj}.  

The rest of the chapter is organized as follows. In section 3.2 we review the corresponding bulk string theory background obtained under the deformation \ref{eq:3.2}. In section 3.3 we compute the entanglement entropy in the deformed dual spacetime theory using the holographic prescription. Following this, we discuss its large and small $L$ limits. In section 3.4 we compute the entropic $c$--function and study its ultraviolet and infrared limits. In section 3.5 we discuss the main results and future research directions. We comment on the entropic $c$--function of a double-trace deformed field theory. We also comment on the case where $F < 0$. In appendix \ref{app_two} we collect some of the intermediate results that are required in section 3.3.

\section{Deformed string background}

We begin with string theory on 
\begin{equation}\label{eq:3.6}
AdS_3 \times S^1 \times {\cal N},
\end{equation} 
with Neveu--Schwartz two--form $B$ field. Where the component ${\cal N}$ is an internal six dimensional compact manifold. Its presence is irrelevant in our discussion. The $S^1$ component gives rise in the boundary conformal field theory to a $U(1)$ current algebra generated by the spacetime currents $J$ and ${\bar J}$ \cite{Giveon:1998ns, Kutasov:1999xu}. 

The bosonic part of the world-sheet theory on $AdS_3 \times S^1$ is described by the action
\begin{equation}\label{eq:3.7}
S = {k\over 2\pi}\int d^2z(\partial \phi{\bar \partial}\phi  + e^{2\phi} \partial {\bar \gamma} {\bar \partial}\gamma + {1\over k}\partial \psi {\bar \partial \psi}),
\end{equation}
where $(\phi, \gamma, {\bar \gamma})$ are the coordinates on $AdS_3$, and $\psi \sim \psi + 2\pi$ is the coordinate on $S^1$. The boundary of $AdS_3$ is located at $\phi = +\infty$. The coordinates $\gamma$ and $\bar \gamma$ are
\begin{equation}\label{eq:3.8}
l_s\gamma = t + x, \quad l_s \bar \gamma = -t + x.
\end{equation}
The level $k$ is given by
\begin{equation}\label{eq:3.9}
l^2 = l_s^2k,
\end{equation}
where $l$ is the radius of curvature of $AdS_3$.

The action has an affine $SL(2, R)_{ L} \times SL(2, R)_{ R} \times U(1)_{ L} \times U(1)_{ R}$ symmetry with left mover world-sheet currents  
\begin{equation}\label{eq:3.10}
J^- = e^{2\phi}\partial\bar\gamma, \quad J^+ = -2\gamma \partial\phi - \partial\gamma + \gamma^2 e^{2\phi}\partial\bar\gamma, \quad J^3 = \gamma e^{2\phi}\partial\bar\gamma - \partial \phi, \quad K = \partial \psi,
\end{equation}
and similar expressions for the right movers ${\bar J}^-, {\bar J}^+$, ${\bar J}^3$ and ${\bar K}$.

Consider deforming the world-sheet theory \ref{eq:3.7} by adding to its Lagrangian the deformation \ref{eq:3.2}. The deformation \ref{eq:3.2} is truly marginal, and therefore, it preserves the conformal symmetry. It breaks the affine $SL(2, R)_{L} \times SL(2, R)_{R} \times U(1)_{L} \times U(1)_{R}$ world-sheet symmetry down to $U(1)_{L}\times U(1)_{R} \times U(1)_{L} \times U(1)_{R}$ affine symmetry. 

The deformation corresponds to a deformation of the metric $g$, dilaton $\Phi$, and Neveu--Schwartz two--form $B$ \cite{Chakraborty:2019mdf, Araujo:2018rho}. We shall refer to the deformed background as ${\cal M}_4$,
\begin{equation}\label{eq:3.11}
ds^2 = d\phi^2 + hd\gamma d{\bar\gamma} + {2h\epsilon_+\over \sqrt{k}}d\psi d{\bar\gamma} + {2h\epsilon_-\over \sqrt{k}}d\psi d\gamma + {1\over k}hf^{-1}d\psi^2,
\end{equation}
\begin{equation}\label{eq:3.12}
e^{2\Phi} = g_s^2e^{-2\phi}h,
\end{equation}
\begin{equation}\label{eq:3.13}
B_{\gamma{\bar\gamma}} = g_{\gamma{\bar\gamma}}, \quad B_{\gamma\psi} = g_{\gamma\psi}, \quad B_{\psi {\bar\gamma}} = g_{{\bar \gamma}\psi},
\end{equation}
where
\begin{equation}\label{eq:3.14}
h^{-1} = e^{-2\phi} + \lambda - 4\epsilon_+\epsilon_-, \quad f^{-1}= h^{-1} + 4\epsilon_+\epsilon_-.
\end{equation}

For $\lambda = 0,\ \epsilon_\pm = 0$ ${\cal M}_4$ reduces to our starting background $AdS_3 \times S^1$. For $\lambda = 0$ ${\cal M}_4$ reduces to a warped $AdS_3 \times S^1$ background \cite{Chakraborty:2018vja, Apolo:2018qpq}. For $\epsilon_\pm  = 0$ ${\cal M}_4$ reduces to a background that is asymptotically $AdS_3 \times S^1$ for large negative $\phi$ and $\mathbb{R}_\phi \times \mathbb{R}^{1, 1}\times S^1$ for large positive $\phi$ \cite{2017arXiv170105576G}. 

It is shown in \cite{Chakraborty:2019mdf} that for a combination of the couplings
\begin{equation}\label{eq:3.15}
\Psi = \lambda - (\epsilon_+ + \epsilon_-)^2, 
\end{equation}
with $\Psi \geq 0$ the geometry is smooth and it has no closed timelike curves. This positivity condition on \ref{eq:3.15} is the dual analogue of the positivity condition on \ref{eq:3.5}.

In this chapter we consider the case in which
\begin{equation}\label{eq:3.16}
\epsilon_+ = {\epsilon\over 2}, \quad \epsilon_- = {\epsilon\over 2}, \quad \Psi \geq 0.
\end{equation}
In this case the background ${\cal M}_4$ is
\begin{equation}\label{eq:3.17}
ds^2 = \alpha'd\phi^2 - hdt^2 + h\left(dx + {\epsilon \sqrt{\alpha' \over k}}d\psi\right)^2 + {\alpha'\over k}d\psi^2, \quad e^{2\Phi} = g_s^2e^{-2\phi}h,  \quad h^{-1} = e^{-2\phi} + \Psi.
\end{equation}
In what follows we work on this background to compute the entanglement entropy and the entropic $c$--function for an interval of length $L$ in the deformed dual conformal field theory using the holographic prescription. We note that for $\lambda < 0$ or equivalently $\Psi < 0$ the function $h$ changes sign beyond a finite radial distance and therefore the metric signature also changes signs. It is not clear in this case how to consistently apply the holographic prescription. We will not consider this case in this chapter. We comment on this later in the discussion section.

\section{Holographic entanglement entropy}

In this section we compute the entanglement entropy in the deformed spacetime conformal field theory for a spatial interval of length $L$ with endpoints at $x = -L/2$ and $x = +L/2$.  It is defined as the Von Neumann entropy,
\begin{equation}\label{eq:3.18}
S = -{\rm Tr} \rho \log \rho,
\end{equation}
corresponding to the reduced density matrix $\rho$ of the subregion $L$. The reduced density matrix $\rho$ is obtained by tracing the global density matrix over the states of the complement of the subregion $L$. The entanglement entropy measures the amount of entanglement between the subregion $L$ and its complement. Due to short range correlations near the boundary of the subregion $L$ the entropy is, however, divergent. To regulate these ultraviolet divergence we introduce a finite cutoff. The entropy is an intrinsic property of the subregion $L$. Thus, it is a useful tool to characterize a field theory along a renormalization group flow. An equally useful and well defined quantity that we will discuss in detail in the next section is the entropic $c$--function which, in any local and Lorentz invariant quantum field theory, is ultraviolet regulator independent and finite. 

In holographic field theories, entanglement entropy is encoded in certain geometrical quantities in the bulk geometry \cite{Ryu:2006ef, Ryu:2006bv, Hubeny:2007xt, Nishioka:2009un, Klebanov:2007ws, Lewkowycz:2013nqa, Dong:2013qoa, Song:2016pwx}. In the context of the AdS/CFT correspondence \cite{Maldacena:1997re, Gubser:1998bc, Witten:1998qj}, entanglement entropy is given by the area of a co--dimension two minimal surface in the bulk geometry  \cite{Ryu:2006ef, Ryu:2006bv, Lewkowycz:2013nqa}. See \cite{Hubeny:2007xt} for a covariant generalization. In this chapter we work under the assumption that the holographic entanglement prescription is also valid for quantum field theories that have dual gravity or string theory descriptions. In what follows we begin by briefly stating this prescription.

Suppose we have a $d$--dimensional holographic quantum field theory. Suppose also the dual string theory is on a background ${\cal M}_{d + 1}$. We assume the background ${\cal M}_{d + 1}$ is static. This is the case in the theory we are considering. To compute the entanglement entropy $S_{\cal R}$ of a given spatial region ${\cal R}$ in the boundary quantum field theory we first find a co--dimension two static surface ${\cal K}$ in the bulk geometry ${\cal M}_{d+1}$ that ends on the boundary of ${\cal R}$. The surface ${\cal K}$ is homologous to ${\cal R}$ and minimizes the area functional. The entanglement entropy $S_{\cal R}$ in the $d$--dimensional boundary quantum field theory is then given by
\begin{equation}\label{eq:3.19}
S_{\cal R} = {{\rm{Area}}({\cal K}) \over 4G_{\rm N}^{(d + 1)}},
\end{equation}
where $G^{(d + 1)}_{\rm N}$ is the $d + 1$--dimensional Newton's constant of the ${\cal M}_{d + 1}$ geometry.

Following the holographic prescription we now compute the entanglement entropy for the spatial interval of length $L$ in the deformation dual to \ref{eq:3.2}. The bulk string geometry \ref{eq:3.17} at a moment of time is  
\begin{equation}\label{eq:3.20}
ds^2 = \alpha' d\phi^2 + hdy^2 + {\alpha'\over k}d\psi^2, \quad h^{-1} = e^{-2\phi} + \Psi,
\end{equation}
where $y = x + {\epsilon\sqrt{\alpha'/k}} \ \psi$.

We now look for a two--dimensional surface $\phi(y, \psi)$ in the geometry \ref{eq:3.20} (and wrapping the internal ${\cal N}$ space) that minimizes globally (in the space of functions) the area functional \ref{eq:3.19} which taking into account the dilaton \cite{Ryu:2006ef, Klebanov:2007ws} is\footnote{Here we rescaled the metric \ref{eq:3.20} by the level $k$.}
\begin{equation}\label{eq:3.21}
S = {\sqrt{k\alpha'}\over 4G_N^{(4)}}\int_0^{2\pi} d\psi \int_{-{L\over 2} + {\epsilon\sqrt{\alpha'/k}} \ \psi}^{+{L\over 2} + {\epsilon\sqrt{\alpha'/k}} \ \psi} dy \ e^{2\phi}\sqrt{{1\over h}\left(1 + {\alpha'\over h} \left(\partial_y\phi\right)^2+ k\left(\partial_\psi\phi\right)^2\right)}, 
\end{equation}
with the boundary conditions
\begin{equation}\label{eq:3.22}
\phi(\pm L/2 + {\epsilon \sqrt{\alpha'/k}} \ \psi, \psi) = \infty, \quad \phi(y, 0) = \phi(y + {\epsilon \sqrt{\alpha'/k}} \ 2\pi, 2\pi),
\end{equation}
where $\psi$ is on $S^1$ that is $\psi \sim \psi + 2\pi$.

We note that under the following continuous and discrete spacetime transformations
\begin{equation}\label{eq:3.23}
\psi \to \psi + \delta, \quad y \to y + {\epsilon \sqrt{\alpha' \over k}}\delta, \quad {\rm{and}} \quad \psi \to -\psi + 2\pi, \quad y \to -y + {\epsilon \sqrt{\alpha' \over k}} 2\pi,
\end{equation}
where $\delta$ is an arbitrary constant, the bulk background \ref{eq:3.20} and the boundary conditions \ref{eq:3.22} are invariant. The surface 
\begin{equation}\label{eq:3.24}
\phi(y, \psi) = \phi(y - {\epsilon\sqrt{\alpha'/k}} \ \psi) = \phi(-y + {\epsilon\sqrt{\alpha'/k}} \ \psi),
\end{equation}
is invariant under the above symmetry transformations \ref{eq:3.23} and thus we expect that it minimizes the area functional. The minimal surface \ref{eq:3.24} is generated by translating the curve, for example at $\psi = 0$, $\phi(y)$, along the line $y = {\epsilon\sqrt{\alpha'/k}} \ \psi$. This curve has the parity symmetry $y \to - y$. The entanglement entropy is then obtained using \ref{eq:3.21}. We find 
\begin{equation}\label{eq:3.25} 
S = {\sqrt{k}\over 4 G_{N}^{(3)}}\int_{-{L\over 2}}^{+{L\over 2}} dx\sqrt{H(U)}\sqrt{1 + \beta(U)(\partial_xU)^2},
\end{equation}
where $G^{(3)}_{\rm N} =  G^{(4)}_{\rm N} / 2\pi l_s$, and 
\begin{equation}\label{eq:3.26}
U = e^{\phi}, \quad U^2h^{-1} = 1 + U^2\left(\lambda - \epsilon^2\right) , \quad U^{-2}H(U) = U^2 h^{-1} , \quad U^4\beta(U) =  (1 + U^2\lambda)\alpha' .
\end{equation}
The boundary conditions \ref{eq:3.22} now take the form 
\begin{equation}\label{eq:3.27}
U(\pm L/2) = U_{\infty},
\end{equation}
where $U_{\infty}$ is an ultraviolet cutoff.

We denote the value at which the curve $U$ takes its minimum value by $U_0$. This value is related to the length of the interval $L$.  This follows from the Euler's variational equation of the action \ref{eq:3.25} with the boundary conditions \ref{eq:3.27}. One finds
\begin{equation}\label{eq:3.28}
L(U_0) = 2\sqrt{H(U_0)}\int_{U_0}^{U_\infty} dU{\sqrt{\beta(U)}\over\sqrt{H(U) - H(U_0)}}.
\end{equation}
The entropy using the Euler's equation of motion becomes 
\begin{equation}\label{eq:3.29}
S = {\sqrt{k}\over  2G_{\rm N}^{(3)}}\int_{U_0}^{U_\infty} dU \sqrt{\beta(U)\over H(U) - H(U_0)}H(U).
\end{equation}

We rewrite the expression \ref{eq:3.28} of the interval length $L$ in terms of the minimum value $U_0$ as
\begin{equation}\label{eq:3.30}
L(U_0) =  {\sqrt{\alpha'}\over U_0} \int_1^{x_\infty}{dx\over x}\sqrt{{(1 + \alpha x)(1 + \alpha_-) \over x(x - 1)( \alpha_- x + \alpha_- + 1) }} , 
\end{equation}
where
\begin{equation}\label{eq:3.31}
\alpha = \lambda U^2_0, \quad \alpha_- = \Psi U^2_0, \quad x_\infty = {U_\infty^2\over U_0^2}.
\end{equation}

The integral \ref{eq:3.30} is ultraviolet convergent and it solves to 
\begin{equation}\label{eq:3.32}
{L \over 2\sqrt{\alpha'\lambda}}= \sqrt{1 + \alpha \over \alpha} E\left(\arcsin\sqrt{{1 + \alpha_- \over 1 + 2\alpha_-}}, \sqrt{1 + 2\alpha_- \over (1 + \alpha)(1 + \alpha_-)}\right),
\end{equation}
where $E(\varphi, k)$ is the incomplete elliptic integral of the second kind\footnote{Here and in what follows please note our notation of elliptic integrals. There are different notations of elliptic integrals in the literature.},
\begin{equation}\label{eq:3.33}
E(\varphi, k) = \int_0^{\varphi} d\theta \sqrt{1 - k^2 \sin^2\theta}.
\end{equation}

In the limit in which $\Psi \to 0^+$ or, equivalently $\alpha_- \to 0^+$, we note that the interval $L$ \ref{eq:3.32} takes the following simpler form
\begin{equation}\label{eq:3.34}
{L \over 2\sqrt{\alpha'\lambda}}= \sqrt{1 + \alpha \over \alpha} E\left(\sqrt{1 \over 1 + \alpha}\right),
\end{equation}
where $E(k) = E(\pi/2, k)$ is the complete elliptic integral of the second kind. Sending $\alpha \to 0^+$ in \ref{eq:3.34} yields
\begin{equation}\label{eq:3.35}
{L \over \sqrt{\alpha'}}= {2\over U_0}.
\end{equation}

We rewrite the entanglement entropy \ref{eq:3.29} as
\begin{equation}\label{eq:3.36}
S = {\sqrt{k\alpha'}\over  4G_{\rm N}^{(3)}}\int_1^{x_\infty}dx\sqrt{{ \alpha x + 1 \over x(x - 1)( \alpha_-x + \alpha_- + 1)  }} \cdot (\alpha_-x + 1).
\end{equation}
We note that for $\alpha_- \neq 0$ the entropy diverges as $S \sim x_\infty$, and in the limit in which we take $\alpha_- \to 0^+$ with $\alpha \neq 0$ it diverges as $S \sim \sqrt{x_\infty}$. We also note that in the case in which we take both $\alpha \to 0^+, \ \alpha_- \to 0^+$ it diverges as $S \sim {\rm{Log}}(x_\infty)$. The integral \ref{eq:3.36} solves with the ultraviolet cutoff $x_\infty$ to 
\begin{eqnarray}\label{eq:3.37}
  & S = {\sqrt{k\alpha'}\over  2G_{\rm N}^{(3)}} {1\over\sqrt{ (\alpha + 1)(\alpha_- + 1)}}\left\{ \left({ \alpha_-} + { \alpha}  - \alpha\alpha_-{d\over d\xi} \right)\right.\cr
   &\left.\left[{1 \over \xi + 1} \cdot \Pi \left(\arcsin\sqrt{{\alpha_- + 1\over 2\alpha_- + 1}\cdot \left( {1-{1\over x_\infty}}\right)}, {2\alpha_- + 1 \over (\xi + 1)( \alpha_- + 1)}, \sqrt{{2\alpha_- + 1 \over (\alpha + 1)( \alpha_- + 1)}}\right)\right]_{\xi = 0} \right.\cr 
& \left. + \ F\left(\arcsin\sqrt{{\alpha_- + 1\over 2\alpha_- + 1}\cdot \left(1 - {1\over x_\infty}\right) }, \sqrt{{2\alpha_- + 1 \over (\alpha + 1)( \alpha_- + 1)}}\right)\right\},
\end{eqnarray}
where $\Pi(\varphi, n, k)$ is the incomplete elliptic integral of the third kind, and $F(\varphi, k) = \Pi(\varphi, 0, k)$ is the incomplete elliptic integral of the first kind,
\begin{equation}\label{eq:3.38}
\Pi(\varphi, n, k) = \int_0^{\varphi} {d\theta\over (1 - n\sin^2\theta)\sqrt{1 - k^2\sin^2\theta}}.
\end{equation}

In the limit $\Psi \to 0^+$ or, equivalently $\alpha_- \to 0^+$, the entropy \ref{eq:3.37} gives
\begin{eqnarray}\label{eq:3.39}
 S = & { \sqrt{k\alpha'}\sqrt{1 + \alpha }\over  2G_{\rm N}^{(3)}}\left[F\left(\arcsin\sqrt{ 1 - {1\over x_\infty} }, \sqrt{1\over 1 + \alpha}\right) - E\left(\arcsin\sqrt{ 1 - {1\over x_\infty} }, \sqrt{1\over 1 + \alpha}\right)\right] \cr
   &+  { \sqrt{k\alpha'}\over  2G_{\rm N}^{(3)}}\sqrt{(\alpha x_\infty + 1)\cdot \left(1 - {1\over x_\infty}\right) }.
\end{eqnarray}
Taking $\alpha \to 0^+$ in \ref{eq:3.39} gives 
\begin{equation}\label{eq:3.40}
S = { \sqrt{k\alpha'}\over  2G_{\rm N}^{(3)}}{\rm{Log}}(2\sqrt{x_\infty}). 
\end{equation}

In the rest of the current section we study the above results in turns.

\subsection{Case \texorpdfstring{$\Psi = 0: \lambda = 0, \ \epsilon = 0$}{Psi = 0: Lambda = 0, Epsilon = 0}}

In this case we have for the interval length $L$ and entanglement entropy $S$ from \ref{eq:3.35} and \ref{eq:3.40}
\begin{equation}\label{eq:3.41}
{L \over \sqrt{\alpha'}}= {2\over U_0}, \quad S = { \sqrt{k\alpha'}\over  2G_{\rm N}^{(3)}}{\rm{Log}}\left(2{U_\infty\over U_0}\right).
\end{equation}
We write the entropy as
\begin{equation}\label{eq:3.42} 
S = {c\over 3}{\rm{Log}}\left(2 {L\over L_{\Lambda}}\right), \quad {L_{\Lambda}\over \sqrt{\alpha'}} := {2\over U_\infty}, \quad c = { 3\sqrt{k\alpha'}\over  2G_{\rm N}^{(3)}},
\end{equation}
where $L_{\Lambda}$ is an ultraviolet cutoff. This result is the well--known entanglement entropy for a two--dimensional holographic conformal field theory with (Brown--Henneaux) central charge $c$ that is dual to pure $AdS_3$  \cite{Ryu:2006bv, Holzhey:1994we, Calabrese:2009qy}.

\subsection{Case \texorpdfstring{$\Psi = 0: \lambda = \epsilon^2 \neq 0$}{Psi = 0: Lambda = Epsilon squared != 0}} 

In this case we have from \ref{eq:3.34} and \ref{eq:3.39} that the interval length $L$ and entanglement entropy $S$ are given by
\begin{equation}\label{eq:3.43} 
{L \over 2\sqrt{\alpha'\lambda}}= \sqrt{1 + \alpha \over \alpha} E\left(\sqrt{1 \over 1 + \alpha}\right),
\end{equation}
\begin{eqnarray}\label{eq:3.44}
 S = & { \sqrt{k\alpha'}\sqrt{1 + \alpha }\over  2G_{\rm N}^{(3)}}\left[F\left(\arcsin\sqrt{ 1 - {1\over x_\infty} }, \sqrt{1\over 1 + \alpha}\right) - E\left(\arcsin\sqrt{ 1 - {1\over x_\infty} }, \sqrt{1\over 1 + \alpha}\right)\right] \cr
   &+  { \sqrt{k\alpha'}\over  2G_{\rm N}^{(3)}}\sqrt{(\alpha x_\infty + 1)\cdot \left(1 - {1\over x_\infty}\right) }, \quad \alpha = \lambda U_0^2, \quad x_\infty = {U_\infty^2\over U_0^2}.
\end{eqnarray}
Note that this case corresponds taking $F = 0$ in \ref{eq:3.5}.

We find from \ref{eq:3.43} that in the large $U_0$ limit the interval length $L$ asymptotes to a minimum value which we denote by $L_0$. It takes the value
\begin{equation}\label{eq:3.45}
L_0 = \pi \sqrt{\alpha'\lambda} = \sqrt{\pi t}.
\end{equation}
We find using \ref{eq:3.43} the following large $U_0$ expansion of the interval length $L$, 
\begin{equation}\label{eq:3.46}
{L\over L_0} = 1 + {1\over 4}\cdot {1\over \alpha} - {3\over 64} \cdot {1\over \alpha^2} +{\cal O}\left({1\over \alpha^3}\right), \quad \alpha = \lambda U_0^2.
\end{equation}
Inverting the above equation one finds 
\begin{equation}\label{eq:3.47}
\alpha = {1\over 4\xi} -{3\over 16} +{\cal O}(\xi), \quad \xi = {L\over L_0} - 1.
\end{equation}
We will use this result momentarily to write the entropy in the ultraviolet in terms of the length $L$.

The small $U_0$ expansion of the interval length $L$ is
\begin{equation}\label{eq:3.48}
{L\over\sqrt{\alpha'}} = {2\over U_0}\left(1 - {1\over 4}\alpha \ln \alpha + {1\over 32}\alpha^2\ln \alpha + {\cal O}(\alpha^3)  \right) , \quad \alpha = \lambda U_0^2.
\end{equation}
The leading term corresponds to the deep $AdS_3$ geometry (see \ref{eq:3.41}). Therefore, in the large $L$ limit the surface is deep inside the bulk. We note also that the correction starts at order $\lambda = \epsilon^2$. 

Inverting equation \ref{eq:3.48} we find
\begin{equation}\label{eq:3.49}
\alpha = \left({2\over \pi}\cdot {L_0\over L}\right)^2\left[1 - \left({2\over \pi}\cdot {L_0\over L}\right)^2{\rm{Log}}\left({2\over \pi}\cdot {L_0\over L}\right)+{\cal O}\left({2\over \pi}\cdot {L_0\over L}\right)^4 \right].
\end{equation}

We can similarly study the large and small $U_0$ or, equivalently, the large and small $L$ limits of the entanglement entropy \ref{eq:3.44}. One finds in the large interval length $L$ limit
\begin{equation}\label{eq:3.50}
S = { c\over 3}\left[\sqrt{\alpha x_\infty} -{1\over 2}{\rm{Log}}(\alpha) -{\alpha\over 8}{\rm{Log}}(\alpha) +{\cal O}\left(\alpha^2\right)\right], \quad \alpha = \lambda U_0^2, \quad x_\infty = {U_\infty^2\over U_0^2},
\end{equation}
which upon using \ref{eq:3.49} gives
\begin{equation}\label{eq:3.51}
S = { c\over 3}\left[{L_0\over L_{\Lambda}} -{\rm{Log}}\left({2\over \pi} \cdot {L_0\over L}\right) + {1\over 4}\left( {2\over \pi} \cdot {L_0\over L}\right)^2{\rm{Log}}\left({2\over \pi} \cdot  {L_0\over L}\right) + {\cal O}\left({2\over \pi}\cdot {L_0\over L}\right)^4 \right], 
\end{equation}
where
\begin{equation}\label{eq:3.52}
L_{\Lambda} := {\pi\sqrt{\alpha'}\over U_\infty},
\end{equation}
 and $L_{\Lambda}$ is an ultraviolet cutoff. The leading logarithmic term is the contribution from the $AdS_3$ region found deep inside the bulk. The coefficient of this term is $-c/3$, as expected. We note that the logarithmic terms depend on the ratio $L/L_0$, and thence $L_0$ sets the non--locality scale of the theory. This will become even more evident in the next section.

As we approach $L_0$ we find
\begin{equation}\label{eq:3.53}
S = {c\over 3}\left[\sqrt{\alpha x_\infty} +{\pi\over 4}{1\over \alpha^{1/2}} - {\pi\over 32}{1\over \alpha^{3/2}} + {\cal O}\left({1\over \alpha^{5/2}}\right) \right], \quad \alpha = \lambda U_0^2, \quad x_\infty = {U_\infty^2\over U_0^2},
\end{equation}
which simplifies using \ref{eq:3.47} to
\begin{equation}\label{eq:3.54}
S = {c\over 3}\left[ {L_0\over L_{\Lambda}} + {\pi\over 2}\left({L\over L_0} - 1\right)^{{1\over 2}} - {\pi\over 16}\left({L \over L_0} - 1\right)^{{3\over 2}} + {\cal O}\left(\left({L\over L_0} - 1\right)^{5\over 2}\right) \right].
\end{equation}
We note that there is no a logarithmic correction term. The entanglement entropy instead shows a square root area law correction at next--to--leading order. The entanglement entropy scales as the square root of the length of the interval. Such power law scaling is quite peculiar and interesting. In local and Lorentz--invariant even--dimensional quantum field theories, however, in general the presence of a logarithmic correction term is generic, and its coefficient is expected to be universal \cite{Casini:2006hu}. It would be interesting to understand this theory better. We discuss its entropic $c$--function in the next section.

\subsection{Case \texorpdfstring{$\Psi > 0: \epsilon = 0$}{Psi = 0: Epsilon = 0}}

This case is studied in \cite{Chakraborty:2018kpr}. Setting $\alpha = \alpha_-$ in \ref{eq:3.32} and \ref{eq:3.37} we find for the length $L$ and entanglement entropy $S$
\begin{equation}\label{eq:3.55}
{L \over 2\sqrt{\alpha'\lambda}}= \sqrt{1 + \alpha \over \alpha} E\left(\arcsin\sqrt{{1 + \alpha \over 1 + 2\alpha}}, \sqrt{1 + 2\alpha \over 1 + 2\alpha + \alpha^2}\right),
\end{equation}
\begin{eqnarray}\label{eq:3.56}
 & S = {\sqrt{k\alpha'}\over  2G_{\rm N}^{(3)}} {1\over \alpha + 1}\left\{ \left(2{ \alpha}  - \alpha^2{d\over d\xi} \right)\right.\cr
& \left.\left[{1 \over \xi + 1}  \Pi \left(\arcsin\sqrt{{\alpha + 1\over 2\alpha + 1} \left( {1-{1\over x_\infty}}\right)}, {2\alpha + 1 \over (\xi + 1)( \alpha+ 1)}, \sqrt{{2\alpha+ 1 \over\alpha^2 + 2\alpha + 1}}\right)\right]_{\xi = 0} \right.\cr
%&\left.\left[{1 \over \xi + 1} \cdot \Pi \left(\arcsin\sqrt{{\alpha + 1\over 2\alpha + 1}\cdot \left( {1-{1\over x_\infty}}\right)}, {2\alpha + 1 \over (\xi + 1)( \alpha+ 1)}, \sqrt{{2\alpha+ 1 \over\alpha^2 + 2\alpha + 1}}\right)\right]_{\xi = 0} \right.\cr 
&  \left. + \ F\left(\arcsin\sqrt{{\alpha + 1\over 2\alpha + 1} \left(1 - {1\over x_\infty}\right) }, \sqrt{{2\alpha + 1 \over\alpha^2 + 2\alpha + 1}}\right)\right\}, \quad \alpha = \lambda U_0^2, \quad x_\infty = {U_\infty^2\over U_0^2}.
\end{eqnarray}

We find that in the large $U_0$ limit the interval length $L$ asymptotes to a minimum value which we denote by $L_0$ (this should cause no ambiguity)
\begin{equation}\label{eq:3.57}
L_0 = {\pi\sqrt{\alpha'\lambda}\over 2} = {1\over 2}\sqrt{\pi t}.
\end{equation}
We note that there is a factor of 2 difference between \ref{eq:3.45} and \ref{eq:3.57}. We find from \ref{eq:3.55} that the interval length $L$ has the following large $U_0$ expansion
\begin{equation}\label{eq:3.58}
{L\over L_0} = 1+ {2\over  \pi\alpha} + {3\pi - 16\over 16\pi \alpha^2}+ {\cal O}\left( {1\over \alpha^3}\right) , \quad \alpha = \lambda U_0^2.
\end{equation}
Inverting the above equation one finds
\begin{equation}\label{eq:3.59}
\alpha = {2\over \pi \xi} + {3\pi - 16\over 32} + {\cal O}(\xi), \quad \xi = {L\over L_0} - 1.
\end{equation}

The small $U_0$ expansion takes the form 
\begin{equation}\label{eq:3.60}
{L\over\sqrt{\alpha'}} = {2\over U_0}\left[1 - {\alpha^2\over 4}{\rm{Log}}(\alpha) + {3\over 8}\alpha^3 {\rm{Log}}(\alpha) + {\cal O}\left(\alpha^4\right) \right] , \quad \alpha = \lambda U_0^2.
\end{equation}
We note that for a long interval the surface is deep inside the bulk in the $AdS_3$ region. We also note that the term linear in $\alpha$ is zero. Therefore, the correction starts, in this case, at order $\lambda^2$.

Inverting the above equation we find
\begin{equation}\label{eq:3.61}
\alpha = \left({4\over \pi} {L_0\over L}\right)^2\left[1 - \left({4\over \pi} {L_0\over L}\right)^4{\rm{Log}}\left({4\over \pi} {L_0\over L}\right) + {\cal O}\left(\left({4\over \pi} {L_0\over L}\right)^6\right)\right].
\end{equation}

In the large $L$ limit we find that the entanglement entropy $S$ has the following series expansion 
\begin{equation}\label{eq:3.62}
S = {c\over 3}\left[ {1\over 2}\alpha x_\infty  - {1\over 2}{\rm{Log}}(\alpha) - {1\over 4}\alpha + {\cal O}\left(\alpha^2{\rm{Log}}(\alpha)\right)\right],\quad \alpha = \lambda U_0^2, \quad x_\infty = {U_\infty^2\over U_0^2},
\end{equation}
which using \ref{eq:3.61} gives
\begin{equation}\label{eq:3.63}
S = {c\over 3}\left[ {1\over 2}\cdot {L_0^2\over L^2_{\Lambda}}\cdot {16\over \pi^2} - {\rm{Log}}\left({L_0\over L}\cdot {4\over \pi}\right) - {1\over 4}\left({L_0^2\over L^2}\cdot {16\over \pi^2}\right) +  {\cal O}\left({L_0^2\over L^2}\cdot {16\over \pi^2}\right)^2\right], \quad U_\infty := {2\sqrt{\alpha'}\over L_\Lambda}.
\end{equation}
The (leading) logarithmic term is due to the deep $AdS_3$ region in the bulk. The coefficient of this term is $-c/3$, as expected. We also note that the arguments of the leading logarithmic terms in \ref{eq:3.51} and \ref{eq:3.63} are also equal despite the different $L_0$ values.

 As we approach $L_0$ the entropy $S$ takes the form 
\begin{equation}\label{eq:3.64}
S = {c\over 3}\left[{1\over 2}\alpha x_\infty - {1\over 2}{\rm{Log}}(\alpha) + {\cal O}\left({1\over \alpha}\right)\right],\quad \alpha = \lambda U_0^2, \quad x_\infty = {U_\infty^2\over U_0^2}.
\end{equation}
Using \ref{eq:3.59} this gives
\begin{equation}\label{eq:3.65}
S = {c\over 3}\left[{1\over 2}\cdot {16\over \pi^2}\cdot {L^2_0\over L^2_{ \Lambda}} + {1\over 2}{\rm{Log}} \left( {\pi\over 2 }\left({L\over L_0} - 1\right)\right) + {\cal O}\left({L\over L_0} - 1\right)\right].
\end{equation}
In this limit the geometry is a linearly varying dilaton background. We note that in this case we have a logarithmically divergent term as opposed to the former $\Psi = 0$ case. The coefficient of this term is $c/6$. 

\subsection{Case \texorpdfstring{$\Psi > 0: \epsilon \neq 0$}{Psi > 0: Epsilon != 0}}
In the large $U_0$ limit we find from \ref{eq:3.32}  that the interval length approach a minimum value $L_0$
\begin{equation}\label{eq:3.66}
L_0 = {\pi \sqrt{\alpha'\lambda}\over 2} = {1\over 2}\sqrt{\pi t},
\end{equation}
that is determined solely by $\lambda$. 

The length $L$ has the following large $U_0$ expansion
\begin{equation}\label{eq:3.67}
{L\over L_0} = 1 + {2 - \delta^2\over \pi (1 - \delta^2)\alpha} + {3\pi - 16 + 2(4 - \pi)\delta^2 - \pi \delta^4\over 16 \pi (1 - \delta^2)^2\alpha^2} + {\cal O}\left({1\over \alpha^3}\right), \quad \delta^2 = {\epsilon^2\over \lambda}, \quad \alpha = \lambda U_0^2.
\end{equation}
Inverting this we find 
\begin{equation}\label{eq:3.68}
\alpha = {2 - \delta^2\over \pi(1 - \delta^2)\xi} + {3\pi - 16 +2(4 - \pi)\delta^2 - \pi \delta^4\over 32 - 48\delta^2 + 16\delta^4}+{\cal O}(\xi), \quad \xi = {L\over L_0} - 1.
\end{equation}

The small $U_0$ expansion takes the form
\begin{equation}\label{eq:3.69}
{L\over\sqrt{\alpha'}} = {2\over U_0}\left[1 -{\delta^2\over 4}\alpha {\rm{Log}}(\alpha) - {\alpha^2\over 4}(1 + \delta^2 p){\rm{Log}}(\alpha) + {\cal O}(\alpha^3)\right], \quad \delta^2 = {\epsilon^2\over \lambda}, \quad \alpha = \lambda U_0^2,
\end{equation}
here $p$ is a polynomial in $\delta^2$. The leading term corresponds to the deep $AdS_3$ region. We note that in this case the correction starts at order $\epsilon^2$.

Inverting the above equation one finds
\begin{eqnarray}\label{eq:3.70}
\alpha = \left({4\over \pi}{L_0\over L}\right)^2\left[1 - \delta^2 \left({4\over \pi}{L_0\over L}\right)^2 {\rm{Log}}\left({4\over \pi}{L_0\over L}\right) - \left({4\over \pi}{L_0\over L}\right)^4(1 + \delta^2 p){\rm{Log}}\left({4\over \pi}{L_0\over L}\right)\right. & \cr
\left. + {\cal O}\left(\left({4\over \pi}{L_0\over  L}\right)^6\right)\right].&
\end{eqnarray}\iffalse
\begin{equation}\label{eq:3.70}
\alpha = \left({4\over \pi}{L_0\over L}\right)^2\left[1 - \delta^2 \left({4\over \pi}{L_0\over L}\right)^2 {\rm{Log}}\left({4\over \pi}{L_0\over L}\right) - \left({4\over \pi}{L_0\over L}\right)^4(1 + \delta^2 p){\rm{Log}}\left({4\over \pi}{L_0\over L}\right) + {\cal O}\left(\left({4\over \pi}{L_0\over L}\right)^6\right)\right].
\end{equation}\fi

In the large $L$ limit we find the following expansion for the entropy $S$
\begin{equation}\label{eq:3.71}
S = {c\over 3}\left[{1\over 2}\alpha x_\infty \sqrt{1 - \delta^2} -{1\over 2}{\rm{Log}}(\alpha) - {1\over 4}\alpha\left({1\over (1 - \delta^2)^2} + {1\over 2}\delta^2 {\rm{Log}}(\alpha)\right) + {\cal O}\left(\alpha^2\right)\right],
\end{equation}
which upon using \ref{eq:3.70} gives
\begin{eqnarray}\label{eq:3.72}
S = {c\over 3}\left\{\left({4L_0\over \pi L_\Lambda}\right)^2 \sqrt{1 - \delta^2\over 4} -{\rm{Log}}\left({4L_0\over \pi L} \right) -{1\over 4}  \left({4L_0\over \pi L} \right)^2\left[{1\over (1 - \delta^2)^2} - \delta^2 {\rm{Log}}\left({4L_0\over \pi L} \right) \right]\right. & \cr
+ \left.{\cal O} \left(\left({L_0\over L} \right)^4\right)\right\},\ &
\end{eqnarray}\iffalse
\begin{equation}\label{eq:3.72}
S = {c\over 3}\left[\left({4L_0\over \pi L_\Lambda}\right)^2 \sqrt{1 - \delta^2\over 4} -{\rm{Log}}\left({4L_0\over \pi L} \right) -{1\over 4}  \left({4L_0\over \pi L} \right)^2\left[{1\over (1 - \delta^2)^2} - \delta^2 {\rm{Log}}\left({4L_0\over \pi L} \right) \right]+{\cal O} \left(\left({L_0\over L} \right)^4\right)\right],
\end{equation}\fi
where $L_\Lambda$ is defined in \ref{eq:3.63}. The leading logarithmic term as in the previous cases corresponds to the $AdS_3$ region found deep inside the bulk. The coefficient of this term is $-c/3$, as expected. 

As we approach $L_0$ we find 
\begin{equation}\label{eq:3.73}
S = {c\over 3}\left[{1\over 2}\alpha x_\infty - {2-\delta^2\over 4\sqrt{1 - \delta^2}}{\rm{Log}}(\alpha) + {\cal O}\left({1\over \alpha}\right)\right],
\end{equation}
which using \ref{eq:3.68} gives 
\begin{equation}\label{eq:3.74}
S = {c\over 3}\left[{1\over 2}\cdot {16\over \pi^2}\cdot{L^2_0\over L^2_{\Lambda}} \cdot\sqrt{1 - \delta^2} +{2-\delta^2\over 4\sqrt{1 - \delta^2}}{\rm{Log}} \left(\left({L\over L_0} - 1\right)\cdot {\pi(1 - \delta^2)\over 2 - \delta^2}\right) + {\cal O}\left({L\over L_0} - 1\right)\right].
\end{equation}

We note that the leading logarithmic term has a coefficient that depends on $\delta^2$. 

In the next section we study in the above cases the Casin--Huerta entropic $c$--function.

\section{Entropic \texorpdfstring{$c$}{c}--function}

In quantum field theory entanglement entropy is ultraviolet divergent. It requires an ultraviolet cutoff to regularize the divergence. However, in two--dimensional local and Lorentz--invariant quantum field theories the Casin--Huerta entropic $c$--function \cite{Casini:2006es} which is derived from the entanglement entropy as
\begin{equation}\label{eq:3.75}
C := L{\partial S\over \partial L},
\end{equation}
is independent of the ultraviolet cutoff and finite. The interval length $L$ is interpreted as a renormalization group scale. 

The entropic $c$--function is also a monotonic function of $L$, and at fixed points of renormalization group flow it is proportional to the corresponding central charges. For scale invariant theories the entropic $c$--function is constant; it is independent of $L$.

The entropic $c$--function is a useful tool to probe phase transitions. In this section we study the entropic $c$--function for the different cases we studied in the former section. We study in each case its monotonicity as a function of $L$ and its independence of the ultraviolet regulator. We also examine its behavior in the ultraviolet regime.

\subsection{Case \texorpdfstring{$\Psi = 0: \lambda = 0, \ \epsilon = 0$}{Psi = 0: Lambda = 0, Epsilon = 0}}

In this case we have
\begin{equation}\label{eq:3.76}
C = {c\over 3}.
\end{equation}
This is the result for a two--dimensional holographic conformal field theory with central charge $c$ \cite{Casini:2006es}. It is independent of the interval length $L$. The entropic $c$--function is non--negative and constant.

\subsection{Case \texorpdfstring{$\Psi = 0: \lambda = \epsilon^2 \neq 0$}{Psi = 0: Lambda = Epsilon squared != 0}}

In this case we find that the entropic $c$--function $C(\alpha)$ is given by
\begin{equation}\label{eq:3.77}
C = {c\over 3}\sqrt{1 + \alpha}E\left(\sqrt{1\over 1+\alpha}\right), \quad \alpha = \lambda U_0^2.
\end{equation}
We study the small and large $U_0$ limit, or equivalently the large and small $L$ limits of the entropic $c$--function \ref{eq:3.77}.  

In the large $L$ limit we find 
\begin{equation}\label{eq:3.78}
C = {c\over 3}\left(1 + {2\over \pi^2 \xi^2}{\rm{Log}}\left({2\over \pi}\cdot \xi\right) + {\cal O}\left({1\over\xi^4}\right)\right), \quad L :=  \xi L_0, \quad L_0 = \pi \sqrt{\alpha'\lambda} = \sqrt{\pi t}.
\end{equation}
In the small $L$ limit we find 
\begin{equation}\label{eq:3.79}
C = {c\over 3}\left( {\pi\over 4}\cdot {1\over \xi^{1\over 2}} + {5\pi\over 32}\cdot \xi ^{1\over 2}+ {\cal O}(\xi^{3/2})\right), \quad \xi = {L\over L_0} - 1.
\end{equation}

One can think of $L$ as a renormalization group scale. We note that the entropic $c$--function increases as we ascend the renormalization group, and it diverges in the ultraviolet at $L_0$. At short distances it diverges as
\begin{equation}\label{eq:3.80}
C \sim \xi^{-{1\over 2}}, \quad \xi = {L\over L_0} - 1.
\end{equation}
This is the case since in the ultraviolet the theory is not governed by a fixed point. We also note that the entropic c--function is independent of the ultraviolet cutoff that we introduced to regularize the entanglement entropy. The variable $\alpha$ can be expressed using the result \ref{eq:3.43} in terms of the interval length $L$ and the coupling $t$ to write \ref{eq:3.77} as a function of only $L$ and $t$.

\subsection{Case \texorpdfstring{$\Psi > 0: \epsilon = 0$}{Psi > 0: Epsilon = 0}}

This case is studied in \cite{Chakraborty:2018kpr}\footnote{It is also studied in closely related works \cite{Lewkowycz:2019xse, Grieninger:2019zts}.}. In this case we find that the entropic $c$--function $C(\alpha)$ in closed--form is given by
\begin{equation}\label{eq:3.81}
C = {c\over 3}(1 + \alpha)E\left(\arcsin\sqrt{1 + \alpha\over 1 + 2\alpha},\sqrt{1 + 2\alpha \over (1+\alpha)^2}\right), \quad \alpha = \lambda U_0^2.
\end{equation}

Using the results \ref{eq:3.59} and \ref{eq:3.61} for $U_0$ we find in the large $L$ limit
\begin{equation}\label{eq:3.82}
C  =  {c\over 3}\left(1 + {8\over \pi^2 \xi^2} + {\cal O}\left({1\over \pi \xi}\right)^4\right), \quad \xi := {L\over L_0}, \quad L_0 = {\pi\sqrt{\alpha' \lambda}\over 2} = {1\over 2}\sqrt{\pi t},
\end{equation}
and in the small $L$ limit
\begin{equation}\label{eq:3.83}
C  = {c\over 3}\left({1\over 2}\left(1 - {1\over \xi}\right)^{-1} + {\cal O}\left(1 - {1\over \xi}\right)\right), \quad \xi := {L\over L_0}.
\end{equation}

We also note in this case that the entropic $c$--function is non--negative and increasing. At short distances it diverges as
\begin{equation}\label{eq:3.84}
C \sim \xi^{-1}, \quad \xi = {L\over L_0} - 1.
\end{equation}

Our results \ref{eq:3.82} and \ref{eq:3.83} are in agreement with the corresponding analyses in \cite{Chakraborty:2018kpr}. Our result \ref{eq:3.81} gives a non--perturbative answer and it can be written as a function of the interval length $L$ and the coupling $t$. We also note from \ref{eq:3.81} that the entropic c--function is independent of the ultraviolet cutoff.

\subsection{Case \texorpdfstring{$\Psi > 0: \epsilon \neq 0$}{Psi > 0: Epsilon != 0}}

In this case we find that the $c$--function $C(\alpha, \chi)$ is given by
\begin{equation}\label{eq:3.85}
C = {c\over 3}\sqrt{(1 + \alpha)(1 + \alpha\chi)}E\left(\arcsin\sqrt{1 + \alpha\chi\over 1 + 2 \alpha \chi}, \sqrt{1 + 2\alpha\chi\over (1 + \alpha)(1 + \alpha\chi)}\right), 
\end{equation}
where 
\begin{equation}\label{eq:3.86}
\alpha = \lambda U_0^2, \quad \Psi = \lambda \chi.
\end{equation}

In this case the large $L$ limit of the entropic $c$--function takes the form
\begin{equation}\label{eq:3.87}
C  =  {c\over 3}\left(1 + {8\over \pi^2 \xi^2}\left({1\over{\chi^2}} + (1 - \chi ){\rm{Log}}\left({\pi \xi\over 4}\right)\right) + {\cal O}\left({1\over \pi \xi}\right)^4\right), 
\end{equation}
where
\begin{equation}\label{eq:3.88}
\xi := {L\over L_0}, \quad L_0 = {\pi\sqrt{\alpha' \lambda}\over 2} = {1\over 2}\sqrt{\pi t}.
\end{equation}

In the small $L$ limit we find
\begin{equation}\label{eq:3.89}
C  = {c\over 3}\left({1+ \chi\over4\sqrt{\chi}}\left(1 - {1\over \xi}\right)^{-1} + {\cal O}\left(1 - {1\over \xi}\right)\right), \quad \xi := {L\over L_0}.
\end{equation}

In this case also the entropic $c$--function is non--negative, ultraviolet cutoff independent and increasing. In the ultraviolet it diverges as
\begin{equation}\label{eq:3.90}
C \sim \chi^{-{1\over 2}}\cdot \xi^{-1}, \quad \xi = {L\over L_0} - 1.
\end{equation}

We note that setting $\chi = 1$ in \ref{eq:3.85} gives \ref{eq:3.81}, and setting $\chi = 0$ gives \ref{eq:3.77}. At $\alpha = 0$ it gives \ref{eq:3.76}. The entropic $c$--function $C$ \ref{eq:3.85} and the interval length $L$ \ref{eq:3.32} satisfy the curious relation 
\begin{equation}\label{eq:3.91}
{C\over c_0} = {L\over l_0}\cdot \sqrt{1 + \alpha\chi}, \quad l_0 = {2\sqrt{\alpha'}\over U_0}, \quad c_0 = {c\over 3}, \quad \alpha = \lambda U_0^2,  
\end{equation}
where $l_0$ and $c_0$ can be thought of as the interval length $L$ and the entropic $c$--function $C$ at $\lambda = 0$ or in the infrared, respectively.

\section{Discussion}

In this chapter we computed the (Von Neumann) entanglement entropy and the entropic $c$--function for an interval of length $L$. We found that the entropic $c$--function is given by
\begin{equation}\label{eq:3.92}
C = {c\over 3}\sqrt{(1 + \alpha)(1 + \alpha\chi)}E\left(\arcsin\sqrt{1 + \alpha\chi\over 1 + 2 \alpha \chi}, \sqrt{1 + 2\alpha\chi\over (1 + \alpha)(1 + \alpha\chi)}\right), 
\end{equation}
where\footnote{In terms of the field theory side deformation parameters $\delta^2$ is given by ${\pi \mu^2\over 8 t}$, and $\mu = 2\mu_+ = 2\mu_-$. }
\begin{equation}\label{eq:3.93}
\alpha = \lambda U_0^2, \quad \chi = 1 - \delta^2, \quad \delta^2 = {\epsilon^2\over \lambda} = {\pi \mu^2\over 8 t}.
\end{equation}
The variable $\alpha$ is related to the interval length $L$ via
\begin{equation}\label{eq:3.94}
{L \over 2\sqrt{\alpha'\lambda}}= \sqrt{1 + \alpha \over \alpha} E\left(\arcsin\sqrt{{1 + \alpha\chi \over 1 + 2\alpha\chi}}, \sqrt{1 + 2\alpha\chi \over (1 + \alpha)(1 + \alpha\chi)}\right), \quad \alpha'\lambda = {t\over \pi}.
\end{equation}

We found that the entropic c--function is non--negative and ultraviolet cutoff independent. This is required for a theory that is internally consistent and it is a non--trivial consistency check. Therefore, this provides further evidence that the deformed theory is very sound and under control. We also found that along the renormalization group upflow towards the ultraviolet it is non--decreasing. At long distances it is proportional to the central charge of the original conformal field theory. At short distances it diverges. This is the case since in the ultraviolet the deformed theory is not governed by an ultraviolet fixed point. The minimum distance at which the entropic c--function diverges sets the non-locality scale of the theory. 

\begin{figure}[h!]
\begin{center}
\includegraphics[width=14cm,height=8cm]{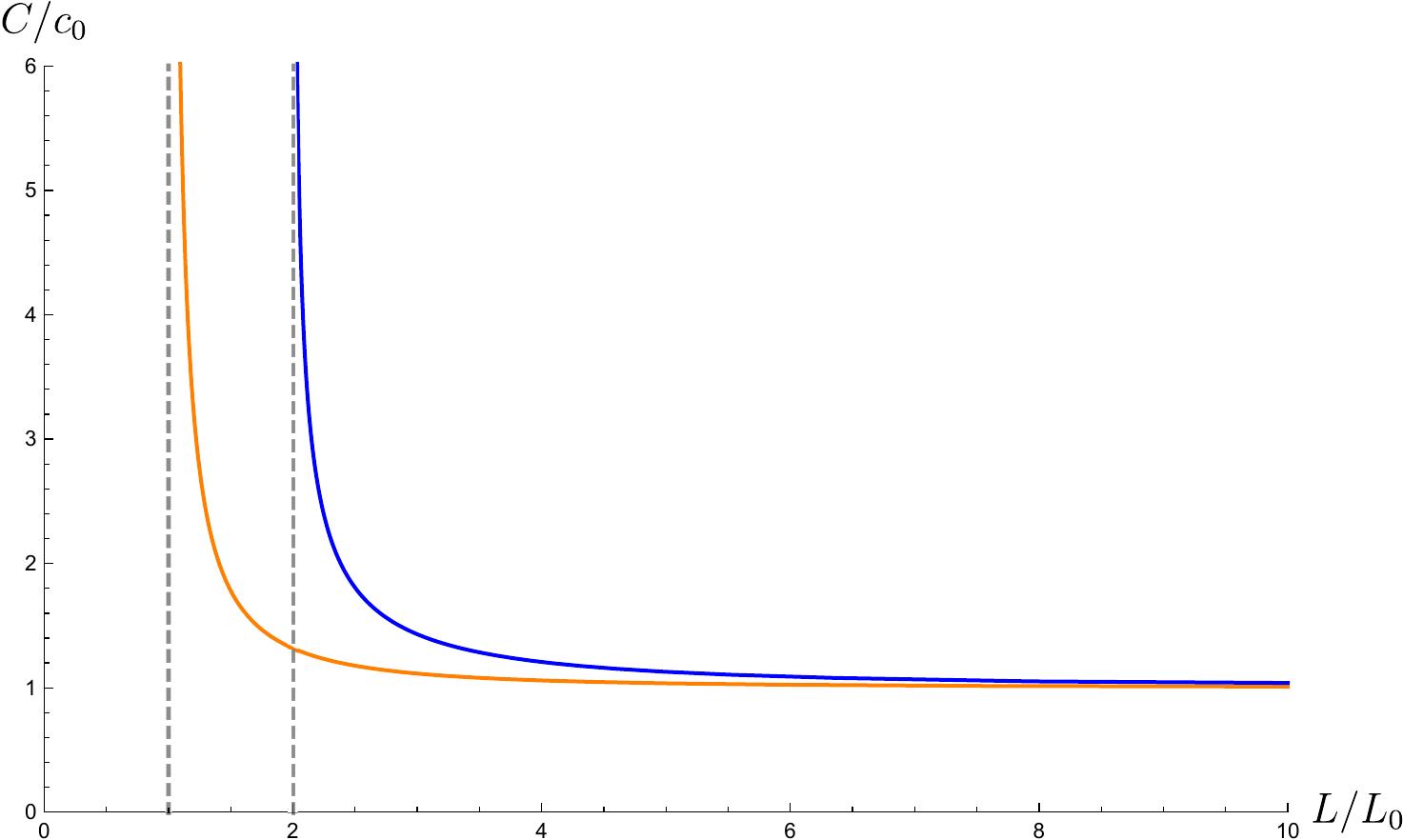}
\caption{Entropic c--function $C$ per $c_0$ as a function of interval length $L$ per $L_0$. Where $c_0 = {c\over 3}$ and $L_0 = {1\over 2}\sqrt{\pi t}$. The orange plot is for $\delta = 0$, and the blue plot is for $\delta^2 = 1$. The (normalized) entropic c--function for the case $\delta = 0$ diverges at $L_0$.}
\label{fig:1}
\end{center}
\end{figure}

In \ref{fig:1} we show plots of the entropic c--function $C$ \ref{eq:3.92} as a function of the interval length $L$ \ref{eq:3.94} for $\delta = 0$ and $\delta^2 = 1$. In the plots we normalized $C$ with respect to $c_0 = {c\over 3}$ and $L$ with respect to $L_0 = {1\over 2}\sqrt{\pi t}$.

In the case in which $F > 0$, the entropic $c$--function diverges in the ultraviolet as
\begin{equation}\label{eq:3.95}
C \sim \chi^{-{1\over 2}}\cdot \xi^{-1}, \quad \xi = {L\over L_0} - 1, \quad L_0 := {\pi\sqrt{\alpha'\lambda}\over 2} = {1\over 2}\sqrt{\pi t}.
\end{equation}
 Note that $L_0$ is determined solely by $\lambda$; it is independent of $\epsilon$.

In the case in which $F = 0$, the entropic $c$--function diverges at short distances as
\begin{equation}\label{eq:3.96}
C \sim \xi^{-{1\over 2}}, \quad \xi = {L\over L_0} - 1, \quad L_0 := \pi \sqrt{\alpha'\lambda} = \sqrt{\pi t}.
\end{equation}
Note that the minimum interval length $L_0$ in this case is twice larger than the corresponding length in the former case. This is depicted in \ref{fig:1} with $\delta = 0$ and $\delta^2 = 1$.

We note that the exponent of $\xi$ in the case in which $F > 0$ is $-1$, and in the case in which $F = 0$ it is $-{1/2}$. This is due to the fact that at short distances the entanglement entropy for the case in which $F > 0$ \ref{eq:3.74} contains a logarithmically divergent term (with a coefficient that depends on $\chi$) whereas in the case in which $F = 0$ \ref{eq:3.54} it does not contain a logarithmically divergent term. In the latter case the entanglement entropy area law exhibits a square root correction \ref{eq:3.54}. The entanglement entropy scales as the square root of the length of the interval,
\begin{equation}\label{eq:3.97}
S = {c\over 3}\left[ {L_0\over L_{\Lambda}} + {\pi\over 2}\left({L\over L_0} - 1\right)^{{1\over 2}} + {\cal O}\left(\left({L\over L_0} - 1\right)^{3\over 2}\right) \right],
\end{equation}
 where $L_\Lambda$ is the ultraviolet cutoff. Such power law scaling is quite distinct and interesting.

 It would be interesting to compute the entropic $c$--function directly in $J{\bar T}, \ T{\bar J}, \ T{\bar T}$ deformed field theory, for example in perturbation theory, and compare it with the bulk calculation result \ref{eq:3.92} (where $\alpha$ is given by \ref{eq:3.94}). It would be also nice to understand better the theory corresponding to the case in which $F = 0$. We leave these for future work.
 
 In this chapter we mainly focused on the case in which $F \geq 0$. As we mentioned in the introduction, in the case $F < 0$ or equivalently $\Psi < 0$, however, the bulk geometry is singular and/or it has either closed timelike curves or no timelike direction. In the field theory side this is related to the presence of states with complex energies. If we look, for example, the case where $\epsilon = 0$, the signature of the boundary (spacetime) metric changes signs as we turn on the coupling $\lambda$. The boundary (spacetime) spatial coordinate $x$ becomes temporal at the outset of the deformation and it is not clear how to consistently apply the holographic prescription in this setting. That is, it is not clear as to whether or not we should consider the region beyond the finite radial distance at which the bulk metric signature changes signs. The holographic prescription may also have to be modified if we excise the region. It would be nice to understand the entropic $c$--function in this case and compare it with results obtained from the closely related double-trace deformations. For example, in the holographic proposal with radial cutoff the entropic $c$--function for an interval of length $L$ in the large $c$ limit is shown to be given by \cite{Lewkowycz:2019xse}\footnote{See also \cite{Grieninger:2019zts} for a related result.}  
 \begin{equation}\label{eq:3.98}
 C = {c\over 3}\cdot {1\over \sqrt{1 - {2\pi \over 3}\cdot {ct\over L^2}}},
 \end{equation}
 with $ct/L^2$ finite. It is also shown in \cite{Lewkowycz:2019xse}\footnote{See also \cite{2018PhRvL.121m1602D} for a related result.} that this result agrees with a field theory calculation performed in a $T{\bar T}$ deformed conformal field theory. We also left out the case $F \geq 0$ and $\mu_+ \neq \mu_-$ which requires using the Hubeny, Rangamani, and Takayanagi prescription \cite{Hubeny:2007xt}. We leave these for future work. 
 
 A fairly similar analysis can be done at finite temperature by considering black hole in the string background \ref{eq:3.11}, \ref{eq:3.12}, \ref{eq:3.13}. At finite temperature the case in which $\epsilon = 0$ is studied in \cite{ Chakraborty:2018kpr, Asrat:2020uib}. It is shown that the inverse maximum (or Hagedorn) temperature which characterizes the theory in the ultraviolet is given by $L_0$ (up to numerical factor of order $0$),
 \begin{equation}\label{eq:3.99}
 \beta_H = 2\sqrt{\pi t}.
 \end{equation}
It is interesting to understand how the maximum temperature depends on the couplings in the other remaining cases. We hope to address this and related questions in future work.

\chapter{Outlook}

A two dimensional CFT that is described by string theory on $AdS_3$ contains the operator $D(x, \bar x)$ that defines the single-trace $T{\bar T}$ deformation. As we mentioned earlier, the long string sector of the string theory is believed to be described in the CFT by a symmetric product theory ${\mathcal S}^p({\mathcal M})$. In the theory ${\mathcal S}^p({\mathcal M})$ the operator $D(x, \bar x)$ is identified with 
\begin{equation}\label{eq:ef1}
D(x, {\bar x}) \equiv \sum_{i = 1}^{p} (T\bar T)_i(x, {\bar x}),
\end{equation}
where the operator $(T\bar T)_i$ is the $T{\bar T}$ operator in the $\it{i}$th copy of the CFT $\mathcal{M}$ which describes a single long string. The integer $p$ is taken to be arbitrarily large. However, string theory also contains a short string sector and in general the boundary CFT that describes both the short and long string sectors may not have a symmetric product structure. Thus, in general we cannot make a similar identification as \ref{eq:ef1}.%, and also it is not known how to independently construct the operator $D$ in the boundary CFT. %This raises the general question that how one would define $D(x, \bar x)$ in a generic quantum field theory that are not holographic.

One of the interesting challenges in the study of single-trace $T\bar T$ deformation is extending it to non-holographic two dimensional (and possibly to higher dimensional) theories that have no symmetric product structures. This requires constructing the operator $D(x, \bar x)$. We also would like to better understand the single-trace deformation in two dimensional holographic theories that cannot be described by symmetric products. That is, we would like to understand how to independently define or interpret the single-trace operator $D(x, \bar x)$ in the boundary (full) CFT.   %This requires constructing the operator $D$ independently in the theories.

%The interpretation that in the bulk the single-trace $T\bar T$ deformation is equivalent to a momentum dependent spectral flow in the sense discussed earlier or equivalently to field dependent coordinate transformations will be useful to understand the operator in the context of the Pohlmeyer reduction.

As a starting point, one way to learn about the operator $D(x, {\bar x})$ and thus the single-trace $T{\bar T}$ deformation in quantum field theories is to exploit in the context of the Pohlmeyer reduction \cite{Pohlmeyer:1975nb} the interpretation that in the bulk the single-trace $T\bar T$ deformation is equivalent to momentum dependent spectral flow in the sense discussed earlier or equivalently to field dependent coordinate transformations in the undeformed AdS background.

Pohlmeyer reduction relates string theory in symmetric spacetimes like AdS to integrable field theories such as the sinh-Gordon theory and its multi-component generalizations. See \cite{Bakas:2016jxp, Hoare:2012nx} for recent reviews. Therefore, Pohlmeyer reduction may shed new light in the study of single-trace deformations in a generic QFT. This will sharpen our understanding on the relations between the single and double-trace deformations. It may also answer some of the questions regrading the symmetric product structure of the full boundary theory.

We hope to have some results in this direction in the future.

\appendix

%\renewcommand\thesection{\TheChapter \Alph{section}}
%\renewcommand\thesection{\Alph{section}}
%\renewcommand\thechapter{\Alph{chapter}}

%\chapter{Appendix}
\chapter{Fourier transform of primary operators in AdS}
%\section{Appendix}
\label{app_one} 

The basic primary operator in $AdS_3$ from which world-sheet vertex operators are constructed is $V_{h, h} := \Phi_h$ (see \cite{Teschner:1997ft, Kutasov:1999xu}), 
\begin{equation}\label{eq:2.85z1q}
 \Phi_h = \frac{1}{\pi}\left(|\gamma - x|^{2} e^{\phi} + e^{-\phi}\right)^{-2h}, 
\end{equation}
here the coordinates $(x, \bar x)$ label the boundary spacetime. The Fourier transform of $\Phi_h(x; z)$ in $(x,\bar{x}) $ is
\begin{equation}\label{eq:2.85z1}
\Phi_h(p; z) = \int d^2x e^{i\vec{p}\cdot\vec{x}}\Phi_h(x; z).
\end{equation}
First we shift $x$ by $\gamma$, and then rescale by $e^{\phi}$, i.e., $x \rightarrow e^{-\phi}(\gamma - x)$.  We write $x := |x|e^{i\theta}$, and now the integral takes the following form
\begin{equation}\label{eq:2.85z2}
\Phi_h(p; z) = \frac{1}{\pi}\cdot e^{i\vec{p}\cdot \vec{\gamma}}\cdot e^{2(h - 1)\phi}\cdot \underbrace{\frac{1}{2}\int dx^2d\theta \left(x^2 + 1\right)^{-2h}e^{i\alpha x \sin\theta}}_{I_1}, \qquad \alpha := pe^{-\phi}.
\end{equation}
Evaluating the $\theta$ integral in $I_{1}$ gives
\begin{equation}\label{eq:2.85z3}
\int_{0}^{2\pi} d\theta e^{i\alpha \sin\theta x} = 2\pi\cdot J_{0}(\alpha x),
\end{equation}
here $J_{\nu}$ is the Bessel function. Now we make use of the definition of gamma function $\Gamma(z)$
\begin{equation}\label{eq:2.85z4}
\Gamma(z) = \int_{0}^{\infty} dt t^{z - 1}e^{-t},
\end{equation}
to write 
\begin{equation}\label{eq:2.85z5}
\left(x^2 + 1\right)^{-2h} = \frac{1}{\Gamma(2h)}\int_{0}^{\infty} dt t^{2h - 1}e^{-(x^{2} + 1)t}.
\end{equation}
Using this, the integral $I_{1}$ becomes 
\begin{equation}\label{eq:2.85z6}
I_1 = \frac{2\pi}{\Gamma(2h)}\cdot \int_{0}^{\infty} dt t^{2h - 1}e^{-t}\underbrace{\int_{0}^{\infty} dx J_{0}(\alpha x)x e^{-x^{2}t}}_{I_2}. 
\end{equation}
Using the following result
\begin{equation}\label{eq:2.85z7}
\int_{0}^{\infty} dt J_{\nu}(\alpha t)e^{-p^2t^2}t^{\nu + 1} = \frac{\alpha^{\nu}}{\left(2p^2\right)^{\nu + 1}}e^{-\frac{\alpha^2}{4p^2}},
\end{equation}
which can be shown using the series expansion of the Bessel function $J_{\nu}$ (see \cite{Korenev}), one finds that
\begin{equation}\label{eq:2.85z8}
I_2 = \frac{e^{-\frac{\alpha^2}{4t}}}{2t},
\end{equation}
and therefore
\begin{equation}\label{eq:2.85z9}
I_1 = \frac{\pi}{\Gamma(2h)}\cdot \int_{0}^{\infty} dt t^{2h - 2}e^{-t - \frac{\alpha^2}{4t}}.
\end{equation}
We have the integral representation of the modified Bessel function of the second kind $K_{\nu}(z)$
\begin{equation}\label{eq:2.85z10}
K_{\nu}(z) = \frac{z^{\nu }}{2^{\nu + 1}}\int_{0}^{\infty} dt t^{-\nu - 1}e^{-t - \frac{z^2}{4t}}.
\end{equation}
We thus note that
\begin{equation}\label{eq:2.85z11}
I_1 =  \frac{\pi}{\Gamma(2h)}\cdot \frac{2^{-2h + 2}}{\alpha^{-2h + 1}}K_{-2h + 1}(\alpha). 
\end{equation}
The Fourier transform $\Phi_h(p; z)$ is then
\begin{equation}\label{eq:2.85z12}
\Phi_h(p; z)=  \frac{2^{-2h + 2}}{\Gamma(2h)}\cdot e^{i\vec{p}\cdot \vec{\gamma}}\cdot e^{2(h - 1)\phi} \cdot \alpha^{2h - 1} \cdot K_{-2h + 1}(\alpha), \qquad\alpha := pe^{-\phi}.
\end{equation}
The modified Bessel function of the second kind $K_{\nu}$ can be expressed in terms of the first kind $I_\nu$
\begin{equation}\label{eq:2.85z13}
K_{\nu}(z) = \frac{\pi}{2} \frac{I_{-\nu}(z) - I_{\nu}(z)}{\sin(\nu \pi)}, \qquad I_{\nu}(z) = \sum_{n = 0}^{\infty} \frac{1}{\Gamma(n + 1)\Gamma(n + \nu + 1)}\left(\frac{z}{2}\right)^{2n + \nu}.
\end{equation}
The first few terms for large positive $\phi$ are
\begin{equation}\label{eq:2.85z14}
\Phi_h(p; z) = \frac{1}{2h - 1}\cdot  e^{i\vec{p}\cdot \vec{\gamma}}\cdot e^{2(h - 1)\phi} + \frac{\Gamma(-2h + 1)}{\Gamma(2h)}\cdot 4^{-2h + 1} \cdot p^{4h - 2}\cdot e^{i\vec{p}\cdot \vec{\gamma}}\cdot e^{-2h\phi}.
\end{equation}
We note that for $h = 1/2$ the two terms are equivalent.

\chapter{Intermediate relevant results}
%\section{Appendix}
\label{app_two} 
We collect the intermediate results required in section 3 to compute the interval length and the Von Neumann entanglement entropy, see \cite{BF}.

We have (with $u > a > b > c > d$),
\begin{equation}\label{eq:a1}
\int_a^u dx{ \sqrt{x - c\over (x - a)(x - b)(x - d)}} = {2\over \sqrt{(a - c)(b - d)}}\left[(b - c)F(\varphi, k) + (a - b)\Pi(\varphi, n, k)\right],
\end{equation}
\begin{equation}\label{eq:a2}
\int_a^u dx{ \sqrt{ x - b\over(x - a)(x - c)(x - d)}} = {2(a - b)\over \sqrt{(a - c)(b - d)}}\Pi(\varphi, n, k),
\end{equation}
\begin{equation}\label{eq:a3}
\int_a^u {dx\over x - b}\sqrt{x - c\over (x - b)(x - a)(x - d)} = {2\over a - b}\sqrt{a - c\over b - d}E(\varphi, k),
\end{equation}
here
\begin{equation}\label{eq:a4}
\varphi = \arcsin\sqrt{(b - d)(u - a)\over (a - d)(u - b)}, \quad n = {a - d\over b - d},\quad k = \sqrt{(b - c)(a - d)\over (a - c)(b - d)}.
\end{equation}
We have (with $a > u \ge b > c, \ r \neq a$),
\begin{equation}\label{eq:a5}
\int^a_u {dx\over x - r}{1\over \sqrt{ (a - x)(x - b)(x - c)}} = {2\over (a - r)\sqrt{a - c}}\Pi(\varphi, n, k),
\end{equation}
here
\begin{equation}\label{eq:a6}
\varphi = \arcsin\sqrt{a - u\over a - b}, \quad n = {a - b\over a - r},\quad k = \sqrt{a - b\over a - c}.
\end{equation}

% Format a LaTeX bibliography
\makebibliography
%\nocite{*}

\providecommand{\href}[2]{#2}\begingroup\raggedright\endgroup

\end{document}